\documentclass[11pt,a4paper]{article}
\usepackage{jcappub}

\usepackage[utf8]{inputenc}
\usepackage{url}
\usepackage{graphicx}
\usepackage{braket}
\usepackage{float}
\usepackage{array}
\usepackage{gensymb}
\usepackage{amsmath}
\usepackage{physics}
\usepackage{amssymb}
\usepackage{textcomp}
\usepackage{dsfont}
\usepackage{amsmath,amsfonts}
\usepackage{amsthm}

\newtheorem*{theorem}{Theorem}

\title{Echoes in multi-ALP scenarios}

\author[a]{Shihabul Haque}
\author[a]{and Sourov Roy}

\affiliation[a]{School of Physical Sciences, Indian Association for the Cultivation of Science, 2A and 2B Raja S.C. Mullick Road, Kolkata 700032, India}

\emailAdd{shihabul1312@gmail.com}
\emailAdd{tpsr@iacs.res.in} 

\abstract{
We present a theoretical study of axion echoes in the context of multiple ALP models. We begin by reviewing the single ALP case, deriving the conditions for resonance 
and echo formation. Starting from a set of $N$ ALPs coupled to the photon, we then derive the relevant echo equations for both coherent and incoherent 
configurations. In the former case, we show that the echo power scales with $N$ leading to sharper amplification and potentially improving sensitivity estimates discussed earlier in
literature. Small mass splittings between the ALPs further increase this amplification, even for a $N=2$ case. In the incoherent scenario, we show that the random phases lead to a suppression of the echo 
power, eventually resulting in observable signals akin to or even weaker than the single ALP case. We also outline the potential experimental implications of our 
results and discuss prospects for detecting these echoes in a wide range of ALP masses.
}


\begin{document}
\maketitle

\section{Introduction}\label{1}
In recent years, the axion has become a point of major interest for beyond-standard-model physics. First proposed as a solution to the Strong CP problem in particle
physics through the Peccei-Quinn mechanism \cite{PQ1, PQ2, PQ3}, the QCD axion \cite{DFSZ1, DFSZ2, KSVZ1, KSVZ2, Khlopov1991, Khlopov1992} grew to encompass another potential role, namely as a dark matter 
(DM) candidate \cite{ALPdm1, ALPdm2, ALPdm3,cline2023}. It was also realised that in several beyond-standard-model contexts, light or ultralight pseudoscalar particles often arise naturally 
and could fulfill the same role as a potential DM candidate. These were dubbed axion-like particles (ALPs). Altogether, the field of axion (or, ALP) physics is rich and dynamic, 
both from a theory perspective as well as a phenomenology perspective. For a detailed description of the current status of the ALP in various contexts, 
see, for example, \cite{Bauer2017,Mimasu2015,Caputo2024,OHare,Marshreview,Sikiviereview,Kar2022,Kar2025,Roy2023,Candon2025}. \\

In the context of ALP phenomenology, a recent area of much interest has been the idea of axion echoes. The most general form of the ALP/photon interaction Lagrangian for a set of
$N$ ALPs is given by, 
\begin{equation}
    \mathcal{L}_{int} = -\frac{1}{4} \sum_{n=1}^N g_{a\gamma\gamma}^n a_n F_{\mu\nu} \tilde{F}^{\mu\nu}
\end{equation}
The form of the above interaction implies two simple processes by which ALPs can interact with photons - one is the standard ALP/photon oscillation wherein the ALP and photon modes mix
with each other in the presence of an external electromagnetic field \cite{raf88}. A second process involves the straightforward decay of an ALP into two photons. With regards to 
this latter situation, it can be shown that the decay rates are amplified in the presence of an external electromagnetic field, i.e., an incident photon, with momentum equal to 
half the mass of the ALP \cite{Echo8}. Because of the conservation of momentum, one of the newly generated photons is emitted in the direction of the initial incident photon while
the other has a momentum opposite to it. Since the latter photon travels opposite to the incident wave, it is termed an ``echo" wave and this entire phenomenon is called an ``axion
echo". The feasibility of detecting such echo waves has been extensively studied in recent literature. In \cite{Arza2019, Echo1, Echo3}, the idea of sending out a beam into space and 
looking for the resulting echo waves, for example by using the 21 CentiMeter Array (21CMA) \cite{Echo2} or the Square Kilometer Array (SKA) \cite{Echo9}, has been discussed in much 
detail. Other studies have focused on echoes generated by photon flux from distinct astrophysical sources. References \cite{Echo5, Echo6} discuss the possibility of looking at echoes 
induced by supernovae remnants which would have had a significant photon flux in the past while \cite{Echo4} considers the flux from Cygnus A, one of the strongest radio sources in 
the sky. Similarly, reference \cite{Sun2023} provides a comprehensive, all-sky analysis of such echo signals based on extragalactic radio point sources, supernova remnants, and
galactic synchotron radiation, also making forecasts for various detectors. Reference \cite{Echo10} discusses axion echoes generated by spheroidal galaxies while reference 
\cite{Echo7} discusses axion echoes in detail, specifically focusing on galactic pulsars. Recently, reference \cite{Sun2025} imposed a constraint on the axion/photon coupling 
strength based on the null detection of such echo signals from the Vela supernova remnant using the Five-hundred-meter Aperture Spherical radio Telescope (FAST).\\

ALPs also arise naturally in several string theory contexts and warped geometry models, several of which often suggest not one but a set of multiple ALPs 
\cite{axiverse, string1, string2}. Such multiple ALPs can also arise from a clockwork mechanism \cite{clockwork1, clockwork2, clockwork3, clockwork4}. Recently, reference 
\cite{Chadha-Day_2024} considered a general ALP anarchy case and reinterpreted previous ALP signals in different experimental contexts. Specifically, they showed that the multiple 
ALP phenomenology could be drastically different from the single ALP case, for example, leading to a weaker signal in experiments such as CAST \cite{CAST}. Similarly, recent references \cite{Lee2025_3, Kondo, Dunsky} also show how the multi-axion situation is distinct from the single ALP scenario and presents novel and interesting 
results (see also \cite{Lee2024_1, Lee2024_2, Higaki2016}). In such a situation, we believe it is worthwhile to explore how the phenomenology of axion echoes might 
change for multiple ALPs and how it can affect future search prospects. \\

In this work, we start from a minimal model and present a theoretical study of axion echoes in the context of multiple ALPs. In section \ref{2}, we 
briefly derive the relevant equations and sensitivity estimates following reference \cite{Arza2019} in the single ALP case. We extend our arguments in section \ref{3} to the multiple ALP case. We 
consider a simple multiple ALP framework with a set of $N$ ALPs, all coupling to the Standard Model (SM) photon, with the masses and coupling strengths distributed in
a certain range, according to some general distribution functions. We consider two broad scenarios, coherent, with the ALP fields oscillating in phase,
and incoherent, when the phases are assumed to be random, and derive the relevant equation in details. For the coherent case, we show that the echo power scales as $N$, leading to 
an amplification depending on the number of ALPs in the theory. We also consider the case of ALPs with small mass splittings as might happen naturally in various string based models 
or in models with ALPs featuring a clockwork mechanism for mass generation and show that the mass splitting parameter causes additional amplification leading to improvements in 
sensitivity estimates even for a minimal $N=2$ case. In section \ref{4}, we present a similar treatment of the incoherent setup and show that the sensitivity estimates are 
always weaker than even the single ALP scenario. Echo calculations are dependent on the exact DM density profile chosen; here, we mostly follow the
standard isothermal profile \cite{Turner} used commonly in axion echo literature in order to facilitate the easy and convenient comparison of results in the two cases.
Our calculations and arguments can be extended to other DM density profiles when required. We present a discussion of our results in section \ref{5}.
Detailed calculations are presented in the appendices.

\section{Perturbative approach for single ALP}\label{2}
In this section, we will briefly review the derivation of the echo wave in the case of a single ALP in the spirit of reference \cite{Arza2019} in order to set the stage for the more complicated case of multiple ALPs. 
The basic idea is as follows - we send out a beam of photons from the earth in the direction of the ALP DM distribution, aiming to capture the resultant echo photons that return back to earth
following the decay of the ALP DM into photons as described earlier. We begin 
with the usual interaction Lagrangian,
\begin{equation}
    \mathcal{L}_{int} = -\frac{1}{4} g_{a\gamma\gamma} a F_{\mu\nu} \tilde{F}^{\mu\nu}
\end{equation}
Our ALP field is denoted by $a(x)$ while the photons are described by $F^{\mu\nu}$. The Maxwell equations in presence of such an interaction term are given by, 
\begin{equation}
    \partial_{\mu} F^{\mu\nu} = g_{a\gamma\gamma} \partial_{\mu}a \tilde{F}^{\mu\nu}
\end{equation}
Since we are considering plane waves, we have, in the Coulomb gauge, 
\begin{equation}
    A^{0} = 0,\ \nabla\cdot\vec{A} = 0
\end{equation}
The ALP dark matter (DM) is usually assumed to be non-relativistic leading to a negligible gradient. This allows us to write, 
\begin{equation}\label{eq5}
    \Box\vec{A}(t, \vec{x}) = - g_{a\gamma\gamma}\partial_{t}a (\nabla \cross \vec{A}(t, \vec{x}))
\end{equation}
The ALP DM, assumed to be static, can be written down as, 
\begin{equation}\label{eq6}
    a(t, \vec{x}) = \mathcal{A}_0 \sin(m_{a}t)
\end{equation}
This is derived from the equations of motion for the ALP field (ignoring backreaction). Here, $\mathcal{A}_0$ represents the amplitude factor, which is related to the DM density as, 
\begin{equation}
    \rho = \frac{1}{2} m^2_a \mathcal{A}_0^2
\end{equation}
Here, $\rho$ is the DM density while $m_a$ is the mass of the ALP. Using eqs. \eqref{eq5} and \eqref{eq6}, we get, 
\begin{equation}
    \Box\vec{A}(t, \vec{x}) = - g \cos(m_{a} t) (\nabla \cross \vec{A}(t, \vec{x}))
\end{equation}
Where, $g = g_{a\gamma\gamma} m_{a} \mathcal{A}_0$. In Fourier space, our equation becomes, 
\begin{equation}
    (\partial_{t}^{2} + p^{2})\vec{A}_{p}(t, \vec{p}) = - i g \cos(m_{a}t) [\vec{p} \cross \vec{A}_{p}(t, \vec{p})]
\end{equation}
This is what we shall be working with.

\subsection{Perturbative solution}
We assume that the photon field can be written as, 
\begin{equation}
    \vec{A}_{p} = \vec{A}_{0}^{p} + \vec{A}_{1}^{p}
\end{equation}
Here, $\vec{A}_{0}^{p}$ is the initial incident radiation while $\vec{A}_{1}^{p}$ is the smaller correction generated by the interaction term with the ALP. Assuming that $g$ is small, 
we have, in the zeroth order,
\begin{equation}
    (\partial_{t}^{2} + p^{2})\vec{A}_{0}^{p} = 0 \Rightarrow \vec{A}_{0}^{p} = \vec{A} e^{-i p t} + \vec{B} e^{i p t}
\end{equation} 
Note that we use $p^\mu = (p,\ \vec{p})$ to denote the 4-momentum of the photon since $p^\mu p_\mu = 0$. Since we start with an outgoing beam of radiation, we must have, 
\begin{equation}
    \vec{A}_{0}(t, \vec{x}) = \vec{a}_{0} e^{i(\vec{k}\cdot\vec{x} - k t)},\ \Dot{\vec{A}}_{0}(t, \vec{x}) = -i k \vec{a}_{0} e^{i(\vec{k}\cdot\vec{x} - k t)}
\end{equation}
This implies,
\begin{equation}
    \vec{A} = \frac{1}{2}\vec{a}_{0}\delta^{(3)}(\vec{k}-\vec{p})\left[1+\frac{k}{p}\right],\ \vec{B} = \frac{1}{2}\vec{a}_{0}\delta^{(3)}(\vec{k}-\vec{p})\left[1-\frac{k}{p}\right]
\end{equation}
Thus, 
\begin{equation}
    \vec{A}_{0}^{p} = \vec{a}_{0}\delta^{(3)}(\vec{k}-\vec{p})\left[\cos(pt) - i \frac{k}{p}\sin(pt)\right] = 
    \vec{a}_{0}\delta^{(3)}(\vec{k}-\vec{p}) e^{-ipt}
\end{equation}
The last equality is imposed by the delta function and the fact that the photon four-momentum has null norm. Now, in first order, we have,
\begin{equation}
    (\partial_{t}^{2} + p^{2})\vec{A}_{1}^{p} = - i g \cos(m_{a}t) [\vec{p} \cross \vec{A}_{0}^{p}] = - \frac{i g}{2} \delta^{(3)}(\vec{k}-\vec{p})
    (\vec{p}\cross\vec{a}_{0}) \Big(e^{i(m_{a} - p)t} + e^{-i(m_{a} + p)t}\Big)
\end{equation}
Let us define, 
\begin{equation}
   \vec{\mathcal{P}}_{kp} = (\vec{p}\cross\vec{a}_{0})\delta^{(3)}(\vec{k}-\vec{p})
\end{equation}
Physically, this quantity reflects both the conservation of momenta and the fact that the echo wave is polarised perpendicular to the initial photon. Then, 
\begin{equation}\label{eqforcedosc}
    (\partial_{t}^{2} + p^{2})\vec{A}_{1}^{p} = - i \frac{g}{2} \vec{\mathcal{P}}_{kp} \Big(e^{i(m_{a} - p)t} + e^{-i(m_{a} + p)t}\Big)
\end{equation}
This is the equation of a simple forced oscillator. The solution to such an equation can be written as (see appendix \ref{A1}), 
\begin{equation}
    \vec{A}^p_1 =\frac{ig}{2m_a} \vec{\mathcal{P}}_{kp}\left[\frac{e^{i(m_a-p)t}}{m_a -2p} + \frac{e^{-i(m_a+p)t}}{m_a + 2p} -\frac{2m_a e^{ipt}}{(m_a-2p)(m_a+2p)}\right]
\end{equation}
It can be seen clearly from the above equation that resonance occurs if $p = m_{a}/2$, i.e., the photon momentum has to be half the axion mass. 
Let us consider the situation at resonance. We then have $p = m_{a}/2 + \delta$ where $\delta$ is very small. We consider only the dominant terms, 
\begin{equation}\label{eqresonance}
    \vec{A}_{1}^{p} \approx -\frac{ig}{2m_a}\vec{\mathcal{P}}_{kp}\left[\frac{e^{i(p-2\delta)t}}{2\delta} - \frac{e^{ipt}}{2\delta}\right] 
    = \frac{ig}{8p\delta}\vec{\mathcal{P}}_{kp}e^{ipt}\left(1 - e^{-2i\delta t}\right)
\end{equation}
We expand the exponential and take the $\delta \rightarrow 0$ limit which gives us,
\begin{equation}
    \vec{A}_{1}^{p} \approx -\frac{gt}{4p}\vec{\mathcal{P}}_{kp} e^{i p t}
\end{equation}
Performing an inverse Fourier transform to go back to real space, we have, 
\begin{equation}
    \vec{A}_{1}(t, \vec{x}) = - \int \frac{d^{3}p}{(2\pi)^{3}}\frac{gt}{4p} 
    \vec{\mathcal{P}}_{kp} e^{i(pt + \vec{p}\cdot\vec{x})}
     = - \frac{g}{4} t (\hat{k} \cross \vec{a}_{0}) e^{i(\vec{k}\cdot\vec{x}+kt)}
\end{equation}
The total solution (up to first order) is then, 
\begin{equation}
    \vec{A}(t, \vec{x}) = \vec{a}_{0} e^{i(\vec{k}\cdot\vec{x}-kt)} - \frac{g}{4} t (\hat{k} \cross \vec{a}_{0}) e^{i(\vec{k}\cdot\vec{x}+kt)}
\end{equation}
The second term arising from the interaction has two distinct features - one, its polarisation is orthogonal to that of the outgoing beam, and, second, it 
travels in the opposite direction to the outgoing wave. This is the \emph{echo wave}. 

\subsection{Power carried by the echo wave}
We now present a brief derivation of the power carried by the echo wave. We start from eq. \eqref{eqresonance}, 
\begin{equation}
    \vec{A}_{1}^{p} = \frac{ig}{8p\delta}\vec{\mathcal{P}}_{kp} e^{i p t} (1 - e^{-2i\delta t}) = 
    - \frac{g}{4p} \vec{\mathcal{P}}_{kp} e^{i p t} \frac{\sin(\delta t)}{\delta} e^{-i\delta t}
\end{equation}
The power is defined as, 
\begin{equation}
    P = \int dp \ |\vec{A}(t, \vec{p})|^2 = \int dp \ \frac{g^2}{16 p^2} |\vec{\mathcal{P}}_{kp}|^2 \Bigg(\frac{\sin(\delta t)}{\delta}\Bigg)^2
\end{equation}
For long times, we have, 
\begin{equation}
    \Bigg(\frac{\sin(\delta t)}{\delta}\Bigg)^2 \rightarrow \pi t \delta(\delta)
\end{equation}
Thus, for the echo wave, 
\begin{equation}
    P  = \Bigg[\frac{\pi g^2 t}{16} \frac{dP_0}{dp}\Bigg]_{p=m_a/2}
\end{equation}
Where, 
\begin{equation}
    \frac{dP_0}{dp} = |\vec{A}^p_0(t, \vec{p})|^2 = a_0^2 \delta^{(3)}(\vec{k}-\vec{p})
\end{equation}
Using $g^2 = (g_{a\gamma\gamma} m_a \mathcal{A}_0)^2 = 2 g^2_{a\gamma\gamma} \rho$ and expressing $\omega/2\pi= \nu$, we finally arrive at, 
\begin{equation}\label{singlepower}
    P = g_{a\gamma\gamma}^2\frac{t}{16} \rho \frac{dP_0}{d\nu}\Big|_{k=m_a/2}
\end{equation}

\begin{figure}[h]
    \centering
        \includegraphics[width=0.85\textwidth]{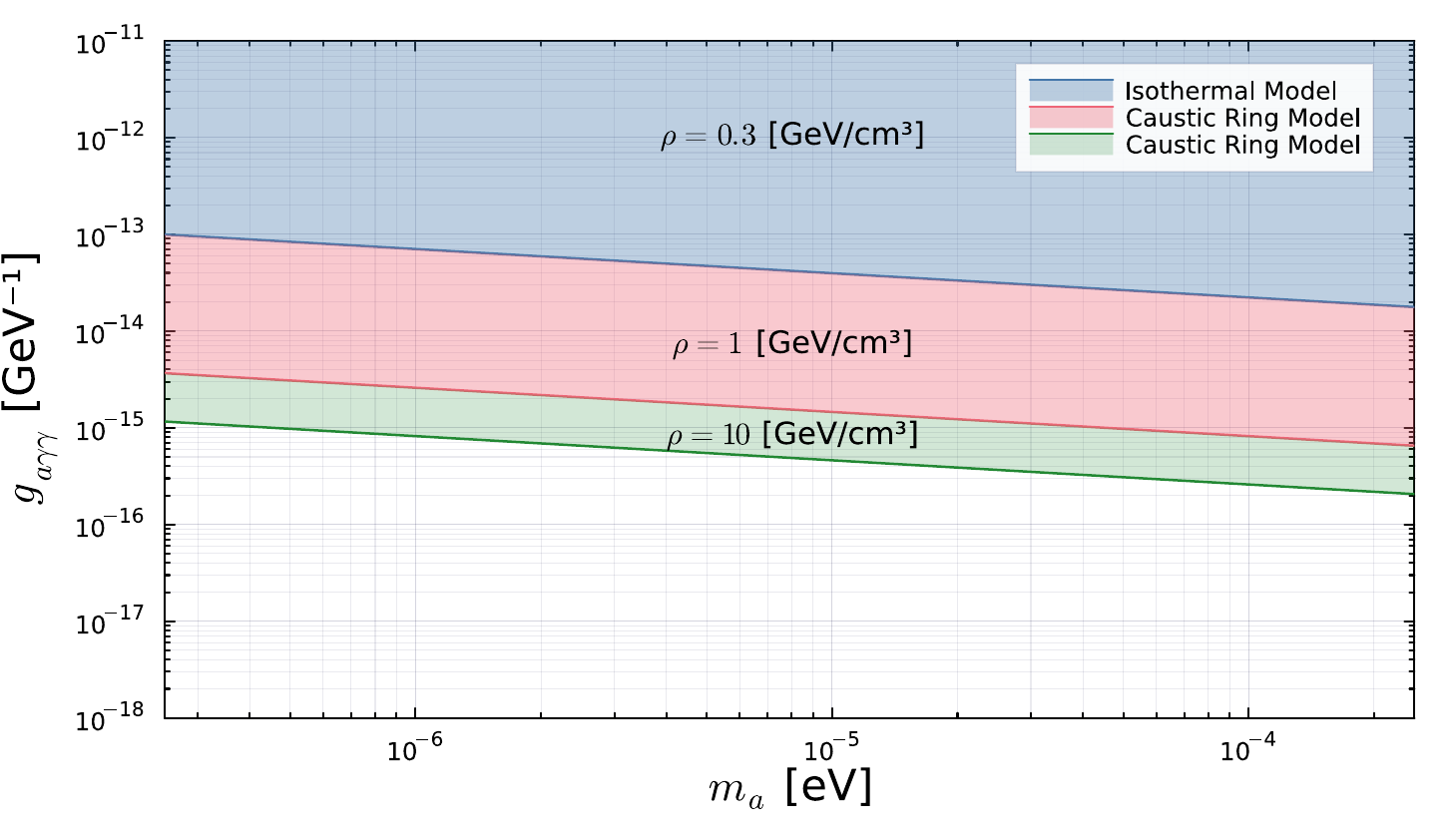}
     \caption{Constraining the parameter space with axion echoes with an outgoing energy of $10$ MW per year per factor of two in ALP mass.}
         \label{Sikivie_Bounds}
\end{figure}
The sensitivity estimates obtained thus are shown in figure \ref{Sikivie_Bounds} for two starkly different DM density profiles, the caustic ring profile and the isothermal profile. The isothermal model
usually proposes a local DM density of around $0.3$ [GeV/cm$^3$] while in the caustic ring model, this value is generally higher because of the earth's proximity to a
caustic ring in the Milky Way galaxy \cite{Duffy, caustic2} leading to much stronger bounds. As mentioned earlier, the nature of the bounds depends on the DM density profile under consideration. For the sake of easy comparison between different cases, we henceforth only consider
the standard isothermal model and follow the observational framework outlined in reference \cite{Arza2019}.\\

For rough experimental estimates, we replace $t$ in eq. \eqref{singlepower} with $C R/v_{\perp}$ where $R$ is the detector radius, $v_{\perp}$ is the magnitude of the velocity of the ALP 
DM perpendicular to the outgoing photon wave and $C$ is an order one number that captures the effect of the configuration of the detector with respect to the DM distribution. 
While our calculations apply to the situation where both the source of photons and the ALP DM are at rest, the results are not altered significantly if we assume that the ALP DM 
has a small velocity with respect to the source in some direction. In this more realistic case, due to the ALP DM velocity, the echo wave is displaced in a transverse direction 
by an amount depending directly on the transverse velocity of the ALP DM. The maximum possible displacement that the echo wave can undergo while still being detectable is simply 
$R$, the radius of the detector, and the time corresponding to this extremal scenario is $R/v_{\perp}$. Therefore, this substitution effectively captures the maximum 
displacement the echo can undergo while still being in the range of our detector and corresponds to the time it takes for the echo wave to travel the distance between our detector 
and the ALP distribution. With this, we have, 
\begin{equation}
P_{c} = g_{a\gamma\gamma}^2 \frac{\rho}{16} \left(C \frac{R}{v_{\perp}}\right)\frac{dP_0}{d\nu}\Big|_{k=m_a/2} 
\end{equation}
This power can be related to the signal-to-noise ratio ($s/n$) of the detector using Dicke's radiometer equation, 
\begin{equation}
\frac{s}{n} = \frac{P_{c}}{T_n} \sqrt{\frac{t_m}{B}}
\end{equation}
\begin{table}[t]
\centering
\caption{Experimental and astrophysical parameters used for sensitivity estimates following \cite{Arza2019}.}
\label{tab:benchmarks}
\begin{tabular}{l c c}
\hline
\hline
Parameter & Value & Additional details \\
\hline
Outgoing beam power, $P_0$ & 10 MW & \\
Detector radius, $R$ & 50 m &  \\
$C$ & 0.30 & Relative configuration between detector and source \\
Noise temperature, $T_n$ & 20 K & System noise \\
\hline
\textbf{Isothermal Model} & & \\
$\rho_{DM}$ & $0.3 \text{ GeV/cm}^3$ & Standard halo density \cite{Turner} \\
$\langle 1/v_{\perp} \rangle$ & $1/124 \text{ km/s}$ & Gaussian average \cite{Arza2019} \\
$B$ & $2.1 \times 10^{-3}\nu$ & \\
\hline
\textbf{Caustic Ring Model} & & \\
$\rho_{DM}$ & $10 \text{ GeV/cm}^3$ & Big flow density \cite{Duffy} \\
$v_{\perp}$ & 5 km/s & Lower bound \cite{Arza2019} \\
$B$ & $ 5 \times 10^{-7} \nu$ & \\
\hline
\hline
\end{tabular}
\end{table}
Here, $T_n$ is the noise temperature of the detector, $t_m$ is the integration time and $B$ is the bandwidth of the signal. The bandwidth is determined by the velocity dispersion of the
ALP distribution - for the isothermal model, this is given by $B = 4\sigma \nu = 2.1 \times 10^{-3}\nu$ where $\nu$ is the frequency of the outgoing photon beam and $\sigma$ is the 
velocity dispersion of the ALPs defined as $\sigma = \sqrt{\langle |\vec{v}|^2 \rangle/3}$. The system noise temperature depends on various factors including the detector location, frequency of the photons, alongside natural backgrounds like
CMB and galactic radiation. We assume a value of $T_n = 20$ K for our estimates. Further, the integration time, assumed to be the same as the emission time of our
outgoing beam, satisfies $t_m \geq R/2 v_{\perp}$. Since our main purpose in this work is to highlight the difference between the multi-ALP and single ALP scenarios, 
we use the same parameter values for both the DM distribution and the detector as in \cite{Arza2019} unless mentioned otherwise (we refer those interested to eq. ($12$) and the 
discussion following it in the above reference for additional details). We summarise these values in Table \ref{tab:benchmarks}.

\section{The multiple ALP scenario}\label{3}
We consider a situation where we have a set of $N$ ALPs in our theory. In principle, these ALPs can interact with both fermionic and vector fields, but we
only focus on their coupling with photons. In general, we do not expect the mass matrix of these ALPs to be diagonal and there will be intermixing; however, in our
case, this does not matter because we only consider couplings to the photon field. To make this clearer, let us consider the most general case with an off-diagonal mass matrix, 
\begin{equation}
    \mathcal{L}_{alp} = \frac{1}{2} \sum_{n, m = 1}^N\left[\delta^{m n}\partial^\mu a_m \partial_\mu a_n - m_{m n}^2 a_m a_n\right]
    - \frac{1}{4}  \sum_{n=1}^m g_{a\gamma\gamma}^n a_n F_{\mu\nu} \tilde{F}^{\mu\nu}
\end{equation}
In the above, only the first $m$ ALPs couple to the standard model photon; the rest do not and are ``hidden". We now consider a transformation that takes us to the physical basis 
where the mass 
matrix is completely diagonal, 
\begin{align}
    a_n = \text{U}_{nk} \tilde{a}_k &\Rightarrow
    \mathcal{L}_{alp} = \frac{1}{2}\sum_{n=1}^N \left[\partial^\mu \tilde{a}_n \partial_\mu \tilde{a}_n - m_n^2 \tilde{a}_n^2\right] -
    \frac{1}{4} \sum_{n=1}^m \left(\sum_{k=1}^N \text{U}_{n k} g_{a\gamma\gamma}^n\right) \tilde{a}_k F_{\mu\nu} \tilde{F}^{\mu\nu}\nonumber
    \\ &\Rightarrow \mathcal{L}_{alp} = \frac{1}{2}\sum_{k=1}^N \left[\partial^\mu \tilde{a}_k \partial_\mu \tilde{a}_k - m_k^2 \tilde{a}_k^2\right] -
    \frac{1}{4} \sum_{k=1}^N \tilde{g}_{a\gamma\gamma}^k \tilde{a}_k F_{\mu\nu} \tilde{F}^{\mu\nu}
\end{align}
Since the original mass matrix is real and symmetric, the transformation has to be orthogonal. As 
we cannot differentiate between the initial and final coupling strengths ($g_{a\gamma\gamma}^n$ and $\tilde{g}_{a\gamma\gamma}^n$) experimentally, it does not matter whether the 
mass matrix is diagonal or whether it needs to be diagonalised first. We can also see that the ``hidden" ALPs also couple to the SM photon as 
part of the transformed field. Therefore, we only work with a situation where all the ALPs are diagonal and are coupled to the SM photon. 
Following the steps outlined in section \ref{2}, we find the following equation in the momentum space for the correction term,
\begin{equation}
    (\partial_{t}^{2} + p^{2})\vec{A}_{1}^{p} = - i \vec{\mathcal{P}}_{kp} \sum_{n = 1}^{N} \frac{g_n}{2} \Big[e^{i(m_n - p)t} + e^{-i(m_n + p)t}\Big]
\end{equation}
With, 
\begin{equation}
    g_n = g_{a\gamma\gamma}^n m_n \mathcal{A}_0
\end{equation}
And, $\vec{\mathcal{P}}_{kp}$ as defined earlier in the single ALP case. In general, the solution will be a superposition of all the different oscillations, 
\begin{equation}
    \vec{A}_{1}^{p} 
    =\sum_{n=1}^{N} \frac{ig_n}{2m_n} \vec{\mathcal{P}}_{kp}\left[\frac{e^{i(m_n-p)t}}{m_n -2p} + \frac{e^{-i(m_n+p)t}}{m_n + 2p} -\frac{2m_n e^{ipt}}{(m_n-2p)(m_n+2p)}\right]
\end{equation}
In this case, we have $N$ separate resonance conditions - the solution will grow whenever $p = m_n/2$ for any $n \in [1, N]$. If all of the masses and couplings are
distinct, then, only one resonance condition can be satisfied at any given time. Suppose we choose $p = m^j_a/2 = p_j$ for any $j$ in the relevant range - this 
means the $j$-th term in the above solution will dominate over the rest. Therefore, we will have, 
\begin{equation}
    \vec{A}_{1}^{p} \approx -\frac{g_j t}{4p_j}\vec{\mathcal{P}}_{k p} e^{i p_j t}
\end{equation}
The real space solution will take the form, 
\begin{equation}
    \vec{A}(t, \vec{x}) = \vec{a}_{0} e^{i(\vec{k}_j\cdot\vec{x}-k_jt)} - \frac{g_j}{4} t (\hat{k}_j \cross \vec{a}_{0}) e^{i(\vec{k}_j\cdot\vec{x}+k_jt)}
\end{equation}
Observationally, this is similar to the case of only one ALP. These conclusions would apply to physically relevant examples such as Kaluza-Klein towers of ALPs where the fields all
have distinct masses (and, therefore, distinct frequencies and resonance conditions) \cite{Basterogil}. In fact, this could be an interesting way to probe the phenomenology of such theories.  
However, here, we consider two separate
kinds of scenarios - a coherent case and an incoherent case - and specifically focus on the situation where there is not much hierarchy in the ALP masses (i.e., the difference
between the most massive ALP and the least massive ALP is small). In general, the ALP fields can be expressed as, 
\begin{equation}
    a_n(t, \vec{x}) = \mathcal{A}_0 \sin(m_n t + \theta_n),\ n\ =\ 1,\, 2,\, \ldots,\, N
\end{equation}
Here, $\theta_n$ is a random phase factor. In the following section, we consider a case where all ALPs have the same phase (which we set to zero for simplicity). This might occur if all
the ALPs have the same production mechanism, for instance, and are produced in the same region of space and time. This is what we call a ``coherent'' scenario. 
In section \ref{4}, we will consider the incoherent case by explicitly taking into account the random phases.

\subsection{Coherent scenario for general distributions}\label{subs0}

The most general equation we can write down is,
\begin{equation}
    (\partial_{t}^{2} + p^{2})\vec{A}_{1}^{p} = - i \mathcal{A}_0 \vec{\mathcal{P}}_{kp} e^{-ipt}\sum_{n = 1}^{N} g_{a\gamma\gamma}^n m_n \cos(m_n t)
\end{equation}
Note that $\vec{\mathcal{P}}_{kp}$ carries no index - the basic kinematic and optical features of the echo process are unaffected. The wave still has polarisation perpendicular to the initial
photon. Let us now assume that we have $N$ ALPs with distinct masses and couplings all distributed according to some distribution functions, $p(m_a)$ and 
$\tilde{p}(g_{a\gamma\gamma})$, respectively. We assume that each of the ALP masses and coupling strengths are distributed according to these function in the intervals 
$[m_L, m_M]$ and $[g_{a\gamma\gamma}^L, g_{a\gamma\gamma}^M]$, respectively. Further, we assume that the distribution functions are normalized and independent of each other, i.e., the joint probability distribution is separable,
\begin{equation}
    P_{\text{joint}}(m_a, g_{a\gamma\gamma}) = p(m_a) \tilde{p}(g_{a\gamma\gamma}) 
\end{equation}
With these assumptions, our masses and coupling strengths become akin to two independent sets of identically distributed random variables. As described in appendix \ref{A3}, in the 
large $N$ limit, by the Law of Large Numbers, this goes to, 
\begin{equation}
    (\partial_{t}^{2} + p^{2})\vec{A}_{1}^{p} = -i N \mathcal{A}_0 \vec{\mathcal{P}}_{kp} e^{-ipt}\int_{g_{a\gamma\gamma}^L}^{g_{a\gamma\gamma}^M} g_{a\gamma\gamma} \tilde{p}(g_{a\gamma\gamma}) \ d g_{a\gamma\gamma}
    \int_{m_L}^{m_M} m_a \cos(m_a t) p(m_a)\ dm_a
\end{equation}
Note that the coupling distribution, $\tilde{p}(g_{a\gamma\gamma})$, does not affect the dynamics of the system. Only the mass distribution does. Therefore, let us write this as, 
\begin{equation}\label{multipleALPcorr}
    (\partial_{t}^{2} + p^{2})\vec{A}_{1}^{p} = -2 i N \vec{D}_{kp} e^{-ipt} Q(t)
\end{equation}
Here, 
\begin{equation}
    \vec{D}_{kp} = \frac{1}{2} \mathcal{A}_0 \vec{\mathcal{P}}_{kp} \int_{g_{a\gamma\gamma}^L}^{g_{a\gamma\gamma}^M} g_{a\gamma\gamma} \tilde{p}(g_{a\gamma\gamma}) \ d g_{a\gamma\gamma}
\end{equation}
And, 
\begin{equation}
    Q(t) = \int_{m_L}^{m_M} m_a \cos(m_a t) p(m_a)\ dm_a
\end{equation}
In order to proceed further, we assume that the mass splitting, defined as $\epsilon = (m_M-m_L)/2$, is small (i.e., $\epsilon << m_L$). In this case, as derived in 
appendix \ref{A4}, the final real space solution can be shown to be, 
\begin{equation}
    \vec{A}(t, \vec{x}) = \vec{a}_{0} e^{i(\vec{k}\cdot\vec{x}-kt)} - 
    N\frac{\tilde{g}t}{2} (\hat{k} \cross \vec{a}_{0})  \left[a_1 + \frac{i}{2} a_2 \epsilon t- \frac{2}{9}\epsilon^2 t^2\right] e^{i(\vec{k}\cdot\vec{x}+kt)}
\end{equation}
Where, 
\begin{equation}\label{fdefine}
    \tilde{g} = m_L  f(\epsilon) \mathcal{A}_0 \int_{g_{a\gamma\gamma}^L}^{g_{a\gamma\gamma}^M} g_{a\gamma\gamma} \tilde{p}(g_{a\gamma\gamma}) \ d g_{a\gamma\gamma},\
    \phantom{.}f(\epsilon) =p(m_L)\epsilon
\end{equation}
The dimensionless parameters $a_1$ and $a_2$ are defined as follows,
\begin{equation}\label{adefine}
    a_1 = 1 + \epsilon\left(\frac{1}{m_L} + \frac{p'(m_L)}{p(m_L)}\right)+ 
    \frac{2\epsilon^2}{3}\left(\frac{2p'(m_L)}{m_L p(m_L)}+\frac{p''(m_L)}{p(m_L)}\right),\
    a_2 = 1 + \frac{4\epsilon}{3}\left(\frac{1}{m_L} + \frac{p'(m_L)}{p(m_L)}\right)
\end{equation}
The power carried by the echo wave is given by,
\begin{equation}
    P_{N}^\epsilon  
    = 2 N \mathcal{Z}(\epsilon, t) \Big[P_{N=1}\Big]_{g_{a\gamma\gamma}^M}
\end{equation}
Where, 
\begin{equation}
    \mathcal{Z}(\epsilon, t) = f(\epsilon)
    \left[\int_{g_{a\gamma\gamma}^L}^{g_{a\gamma\gamma}^M} \frac{g_{a\gamma\gamma}}{g_{a\gamma\gamma}^M} \tilde{p}(g_{a\gamma\gamma}) \ d g_{a\gamma\gamma}\right]^2
    \left[a_1 + \frac{\epsilon}{m_L}a_2\right]^{-1}\left[a_1^2 +\frac{7}{24}\left(a_2^2-\frac{12}{7}a_1\right)\epsilon^2 t^2\right]
\end{equation}
Here, $P_{N=1}$ refers to the power in the original, single ALP case. We now apply our results to a few specific cases.

\subsection{Variable couplings and all masses equal}\label{subs2}
As a first application of our results, let us consider a case with all the ALPs having the same mass, $m_a$, but variable coupling strengths. As noted previously, the coupling strength 
distribution does not affect the dynamics of the situation, only the mass distribution does. Therefore, the contribution of the variation of all the couplings simply enters as an average
over the distribution, denoted by $\langle g_{a\gamma\gamma} \rangle$. For the mass contribution, note that we cannot simply take the limit $\epsilon \rightarrow 0$,
since this collapses the integral in $Q(t)$ to $0$. We also need to consider the variation of the probability distribution. To see this, consider the Taylor expanded
integral, 
\begin{equation}
    Q(t)\Big|_{m_M = m_L+2\epsilon} = 2 m_L \cos(m_L t) p(m_L) \epsilon\left[1 + \epsilon\left(\frac{1}{m_L} + \frac{p'(m_L)}{p(m_L)}\right) + \cdots\right]
\end{equation}
We require that this simply equal $m_a \cos(m_a t)$ in the required limit. This happens if, 
\begin{equation}
    \lim_{\epsilon\rightarrow 0}\ \epsilon p(m_a) = \frac{1}{2}
\end{equation}
Essentially, while the integral limits do collapse to a single point ($m_L = m_a$ since all the masses are equal), we must make sure that the probability distribution also peaks
in exactly the right manner so that the normalisation condition holds true. Combining everything, we find,
\begin{equation}
   \lim_{\epsilon\rightarrow 0}\text{: } a_1 = 1,\ a_2 = 1,\ \tilde{g} = \frac{1}{2} m_a g_{a\gamma\gamma}\mathcal{A}_0,\ \mathcal{Z}(\epsilon, t) = \frac{1}{2}
\end{equation}
The solution is straightforward now, 
\begin{equation}
    \vec{A}(t, \vec{x}) = \vec{a}_{0} e^{i(\vec{k}\cdot\vec{x}-kt)} - N \frac{m_a \langle g_{a\gamma\gamma} \rangle \mathcal{A}_0}{4} t (\hat{k} \cross \vec{a}_{0}) 
    e^{i(\vec{k}\cdot\vec{x}+kt)}
\end{equation}
\begin{figure}[h]
    \centering
    \includegraphics[width=0.85\textwidth]{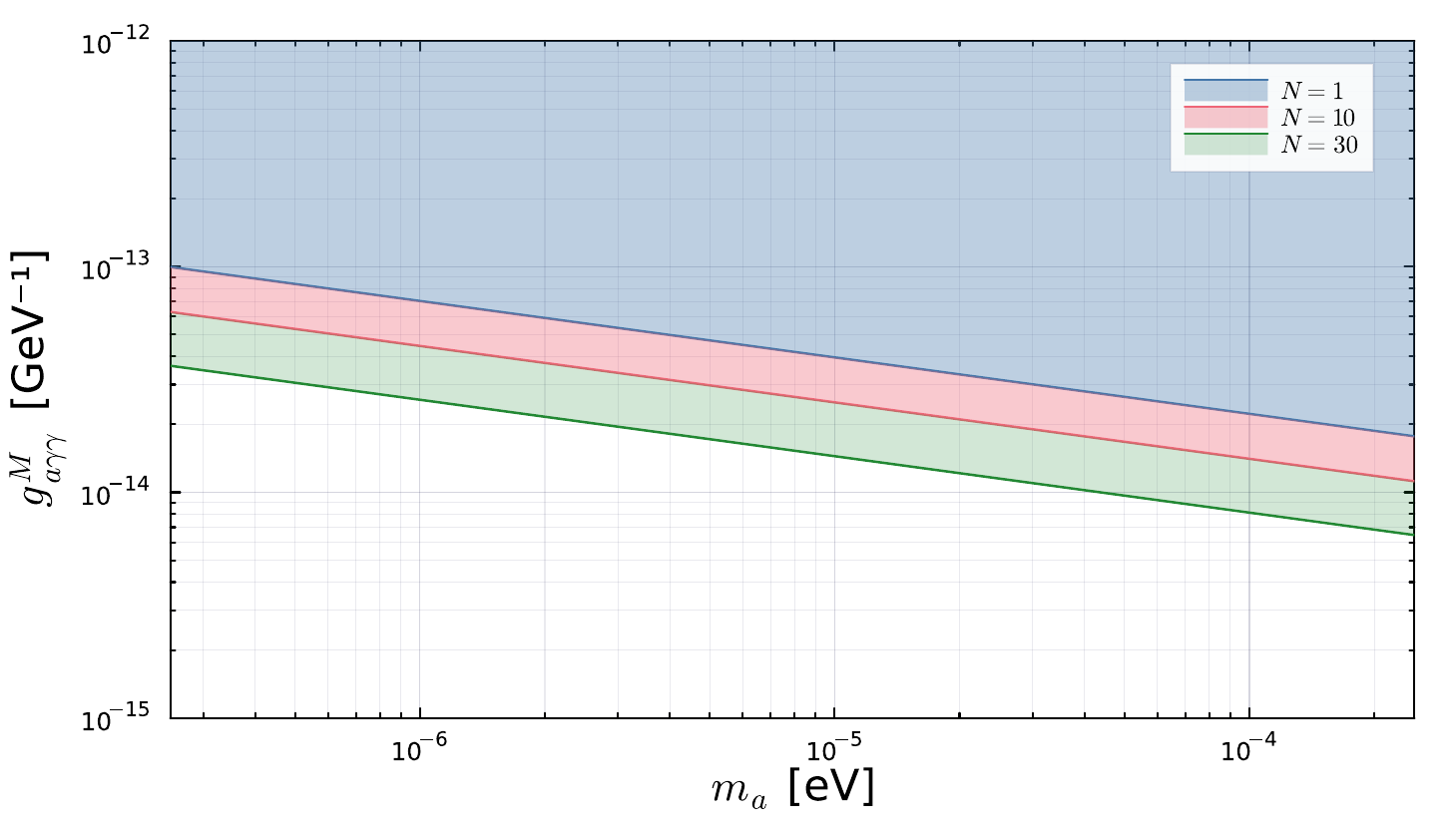}
     \caption{Sensitivity estimates for $N = 10$ and $N = 30$ ALPs with all masses equal compared to the single ALP case for the isothermal model with $\rho = 0.3$ [GeV/cm$^3$].}
         \label{Sikivie_Bounds_10_eqm}
\end{figure}
Clearly, the echo signal is amplified compared to the single ALP case - the strength of the amplification depends on the exact choice of the coupling strengths. To estimate the power, 
we first note that the dark matter energy density is defined as the Hamiltonian density of the ALP sector. Since we ignore the backreaction of the electromagnetic field on the ALP 
fields, this is given by the Hamiltonian density for usual free scalar fields, 
\begin{equation}
    \rho \equiv \mathcal{H}_{\rm ALP} = \sum_{n=1}^{N} \left[\frac{1}{2} (\partial_0 a_n)^2 + \frac{1}{2} (\nabla a_n)^2 
    + \frac{1}{2} m_n^2 a_n^2\right] 
\end{equation}
Since the ALP fields are non-relativistic, we can ignore the spatial gradient term. This gives us,
\begin{equation}
    \rho = \frac{1}{2} \sum_{n=1}^N m_n^2 \mathcal{A}_0^2 = \frac{N}{2} m^2 \mathcal{A}_0^2
\end{equation}
Therefore, the echo wave power can be written down as, 
\begin{equation}
    P_N = N \langle g_{a\gamma\gamma} \rangle^2 \frac{t}{16} \rho \frac{dP_0}{d\nu}\Big|_{k=m_a/2}
\end{equation}
For the purpose of illustration, let us assume that the coupling distribution is uniform over the given range and that $g_{a\gamma\gamma}^M >> g_{a\gamma\gamma}^L$. Then,
\begin{equation}
    P_N = \frac{1}{4} N (g_{a\gamma\gamma}^M)^2 \frac{t}{16} \rho \frac{dP_0}{d\nu}\Big|_{k=m_a/2}
\end{equation}
The equation above is mathematically equivalent to a single ALP case with $g \rightarrow \sqrt{N} g_M/2$. As shown in 
figure \ref{Sikivie_Bounds_10_eqm}, the constraints from the previous case are stronger than the single ALP case. The advantage of our treatment is that even for multiple ALPs with 
differing couplings, we do not need to know the exact distribution of coupling strengths. Starting off with a set of $N$ different coupling strengths, we have derived an answer that
depends on, at most, two free coupling parameters. A general idea of the limits of the parameter space being searched for is sufficient to 
provide a good idea of the kind of signal we expect - this, in turn, can help focus our experimental probes towards the most promising regions of the parameter space.\\
\begin{figure}[h]
    \centering
        \includegraphics[width=0.65\textwidth]{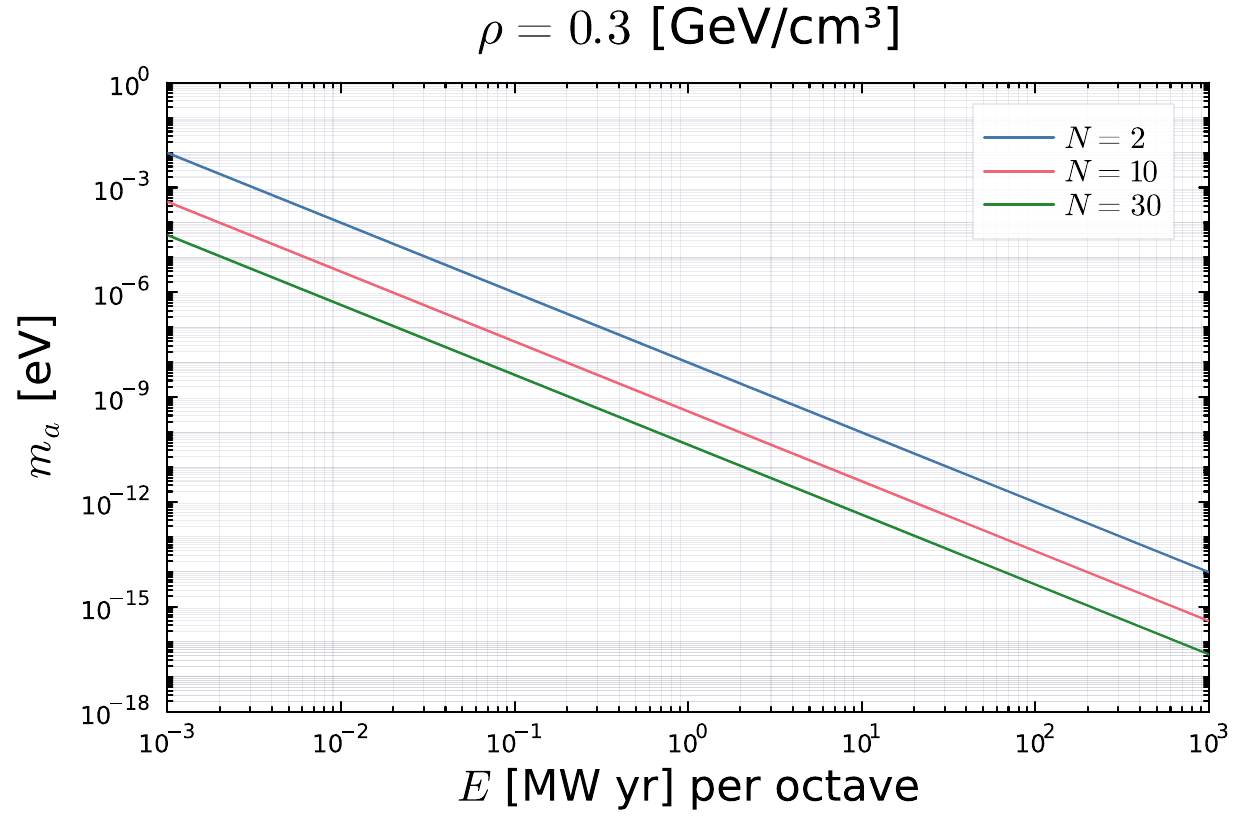}
     \caption{Range of masses that can be probed depending on outgoing wave energy in
     the isothermal model for different $N$.}
        \label{Mass_Range}
\end{figure}

A second interesting application is as follows. If we assume, for example, $g_{a\gamma\gamma}^M \sim 10^{-12}$ [GeV$^{-1}$] based on existing constraints, the above equation gives 
us a range of masses that we can probe depending on the energy of the outgoing photon wave. This is shown in figure \ref{Mass_Range}. This establishes a direct connection between the 
input power and the ALP mass and can serve as a sensitivity estimate - for a given range of ALP masses, we can estimate the rough power needed to produce a detectable signal. As 
before, this is significant especially in such a multi-ALP context where we have shown that an ``effective" coupling strength allows us to determine the features of the 
expected signal without knowing the exact details of the mass and coupling distribution.

\subsection{Variable masses and all couplings equal}\label{subs3}
As a straightforward illustration of our results to a more complicated case, let us consider a situation where the mass distribution is uniform (but narrow), while all of the coupling strengths are same
and equal to $g_{a\gamma\gamma}$ (where $g_{a\gamma\gamma}$ now refers to a specific value for the coupling strength rather than the dummy integration variable as used above). Then, we 
have, 
\begin{equation}
    \tilde{p}(g'_{a\gamma\gamma}) = \delta(g'_{a\gamma\gamma}-g_{a\gamma\gamma}),\ p(m_a) = \frac{1}{m_M - m_L} = \frac{1}{2\epsilon},\ p'(m_a) = 0,\ p''(m_a) = 0
\end{equation}
Then, 
\begin{equation}
    a_1 = 1 + \frac{\epsilon}{m_L},\ a_2 = 1 + \frac{4\epsilon}{3m_L}
\end{equation}
We express our answer in a more dimensionless form by redefining $\epsilon \rightarrow \epsilon m_L$. We find,
\begin{equation}
    \vec{A}(t, \vec{x}) = \vec{a}_{0} e^{i(\vec{k}\cdot\vec{x}-kt)} - 
    N \frac{g_{a\gamma\gamma} m_L \mathcal{A}_0 t}{4} (\hat{k} \cross \vec{a}_{0})
    \left[1 + \epsilon + i \frac{m_L t}{2} \left(1 + \frac{4}{3}\epsilon\right)\epsilon - \frac{2(m_L t)^2}{9}\epsilon^2\right] e^{i(\vec{k}\cdot\vec{x}+kt)}
\end{equation}
\begin{figure}[h]
    \centering
    \includegraphics[width=0.85\textwidth]{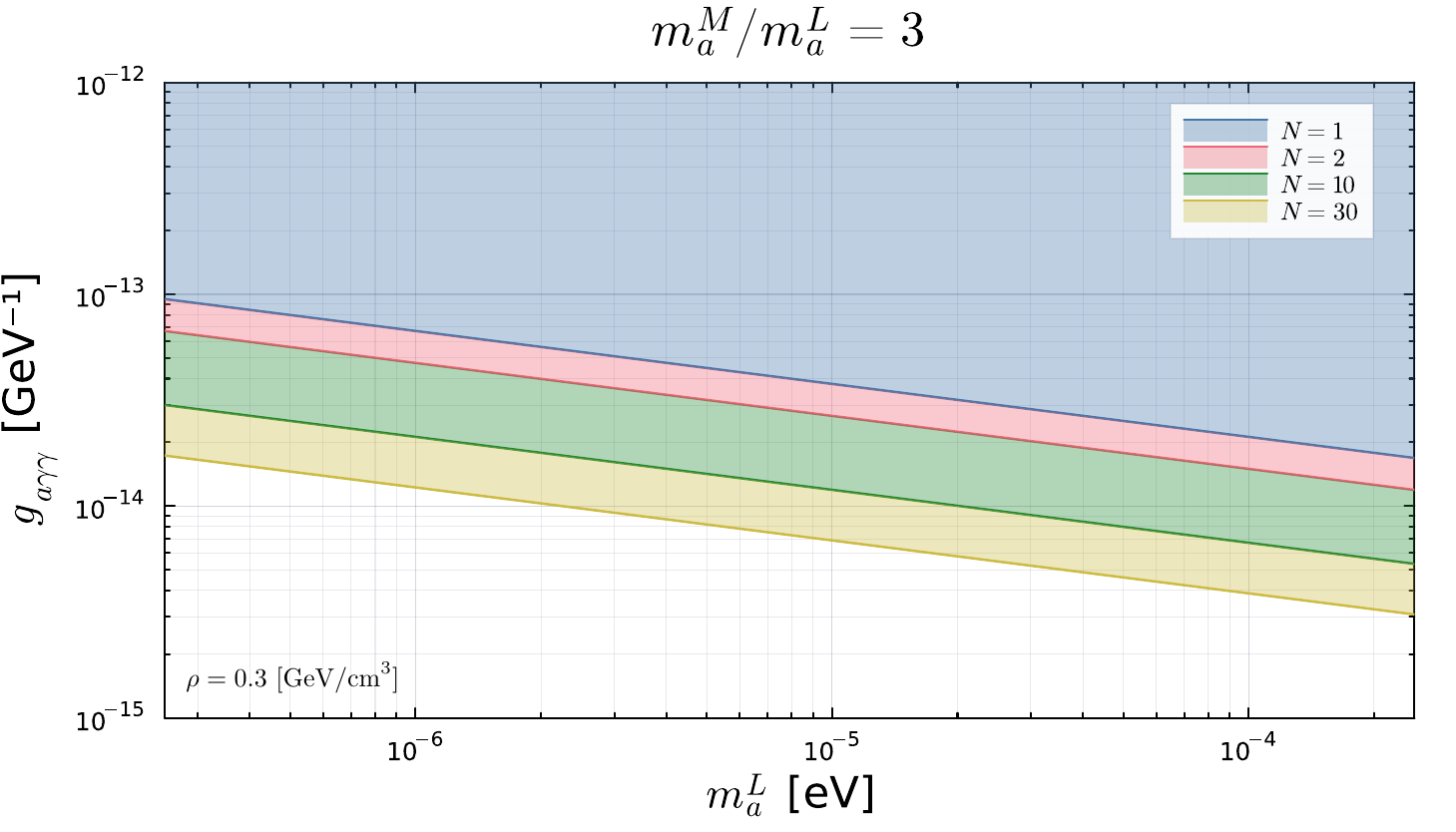}
     \caption{Constraints with $N=2$, $N=10$, $N=30$ and mass ratio $3$ corresponding to $\epsilon = 1$ for the isothermal model with $\rho = 0.3$ [GeV/cm$^3$].}
         \label{Sikivie_Bounds_2_3}
\end{figure}
We also have,
\begin{equation}
    P_N^\epsilon  = N \left[1+2\epsilon+\frac{4}{3}\epsilon^2\right]^{-1} \left[\left(1+\epsilon\right)^2-\frac{5 (m_L t)^2}{24}
    \epsilon^2\left(1-\frac{4}{3}\epsilon-\frac{112}{45}\epsilon^2\right)\right] P_{N=1}
\end{equation}
The echo wave is amplified again in this case. Aside from the usual $N$ amplification, we have an additional 
enhancement depending on the exact difference between the masses and the timescale under consideration. For larger times, the higher degree polynomial terms will dominate with the 
caveat that we will also be approaching the region where the perturbation theory treatment will break down. Technically, the condition for the validity of the perturbative treatment requires, 
\begin{equation}
    \epsilon m_L t = 1 \Rightarrow t = \frac{2}{m_M - m_L}
\end{equation}
In this limit, higher order terms will also contribute, but the coefficients of these terms will get successively smaller, and the major contribution is already captured by the 
terms considered here. For masses on the scale of $10^{-4}$ [eV], for instance, this would amount to a timescale of $t \sim 10^{-11}$ [sec]. This is a short timescale and according to the answer we have derived, it is possible that the echo 
wave is amplified significantly on this timescale. As mentioned earlier, we replace $t$ by $R/v_{\perp}$ in order to arrive at a rough experimental estimate. Since this value is much larger than the 
timescale of amplification ($\sim 10^{-11}$ [sec]), we should be able to detect such echo signals.\\

Consider a situation with $m_L t \sim 1$. Our perturbative answer will be valid for any $\epsilon \leq 1$, in this case. The power will be, 
\begin{align}
    P_N^\epsilon  &= N \left[1+2\epsilon+\frac{4}{3}\epsilon^2\right]^{-1}\left[\left(1+\epsilon\right)^2-\frac{5}{24}\epsilon^2\left(1-\frac{4}{3}\epsilon-\frac{112}{45}\epsilon^2
    \right) + \frac{131}{1620}\epsilon^4\right] P_{N=1}\nonumber\\ &\leq P_N^{\epsilon=1} \approx 1.1 N P_{N=1}
\end{align}
This corresponds to a straightforward additional amplification of $10\%$ aside from the usual $N$ amplification. The sensitivity estimates for this maximal case are plotted in 
figure \ref{Sikivie_Bounds_2_3} for the maximal case with $\epsilon = 1$ for $N=2$, $N=10$ and $N=30$. As we can see, there is an enhancement in the sensitivity estimates even for the 
$N=2$ case compared to the single ALP scenario. As expected, higher $N$ leads to more area of the parameter space being covered.\\

The sensitivity estimates that we show here fall in an area of the parameter space that is mostly unconstrained and can, therefore, be an interesting potential avenue for constraining such
ALPs. The mass range considered here is given by $m_L \in [2.5\times10^{-7}, 2.5\times10^{-3}]$ [eV] aligning with the transparency window of the earth's atmosphere which 
is almost completely transparent to photons in this frequency range (around $0.3 - 30$ [GHz]). This mass range also corresponds to 
QCD axions and therefore, these results can be used to provide stronger or complementary bounds compared to the traditional haloscope based experiments like ADMX \cite{ADMX1,
ADMX2, ADMX3, ADMX4, ADMX5, ADMX6, ADMX7} and CAPP \cite{CAPP1, CAPP2, CAPP3, CAPP4}. However, our results apply, in general, to any mass range (as long as the perturbative condition 
is valid). As such, in principle, our results could also be extended to obtain
newer and complementary bounds on ALPs in lower mass ranges, $10^{-10}$ [eV] $< m_L < 10^{-6}$ [eV], which have been investigated using data from Chandra and Fermi-LAT which are
not ground-based and are, therefore, not constrained to work only in the earth's transparent window.\\

There are also future proposals that would enable us to look at extremely low 
frequency regimes. Two such proposals are NASA's FARSIDE concept \cite{burns2019nasaprobestudyreport,burns2021lunarfarsidelowradio} and the DSL (Discovering the Sky at the Longest 
wavelengths) experiment \cite{DSL} in China, both of which are based on the farside of the moon. The FARSIDE experiment will be designed as an array of low-frequency antennas on the 
lunar farside surface, while the DSL experiment will include a set of satellites in the lunar orbit on the farside. Both these experiments will be designed to detect ultra-low 
frequency signals below $30$ MHz with both aiming for a minimum frequency of around $100$ kHz. For example, the DSL experiment proposes a set of detectors with radius $0.5$ m, this
gives us a lower value of $R/v_{\perp} \sim 10^{-4}$ [sec] for the caustic ring model and $10^{-6}$ [sec] for the isothermal model both of which are still several orders of magnitude
higher than the timescale we require for the perturbative regime to be valid. However, the major issue in such a situation is with the question of sending out photons of such low energy
reliably in a direction of our choosing. For a start, such beams can only be sent out using satellite-based methods because otherwise the earth's atmosphere will not allow them to pass
through. Further, in order to transmit such large wavelengths, the required antennae will also need to be very large. For example, on the earth, very low frequency transmitters use wire 
antennae that go up to several kilometres long, which, in the context of satellites, seems impractical. These issues, however, could be potentially resolved in the future based on
novel technological advancements. Here, we note that while our results can be applied to these scenarios in principle, the practicality of the experimental methods is unclear.

\section{The incoherent scenario}\label{4}
As mentioned earlier, the most general form of the ALP fields includes an additional random phase factor. In this section, we take a look at what happens in such a situation. We
have, 
\begin{equation}
    (\partial_{t}^{2} + p^{2})\vec{A}_{1}^{p} = - i \mathcal{A}_0 \vec{\mathcal{P}}_{kp} e^{-ipt}\sum_{n=1}^{N}g_{a\gamma\gamma}^n m_n 
    \cos(m_n t + \theta_n)
\end{equation}
In general, averaging over the random phase on the RHS makes the drive term vanish. Instead, we focus on replacing the randomised sum with an ``effective'' driving term by using
the root mean square value, similar to how we treat random walks. This difference in treatment underscores the important fact that the two cases - coherent and incoherent - are not
equivalent to each other in any limiting sense. They are separate, distinct scenarios that must be treated separately. As before, we start from the general case and then consider a few
specific cases to illustrate the broad reach of our results. 

\subsection{Incoherent scenario for general distributions}\label{subs0_2}
We start with,
\begin{equation}\label{incoherentoriginal}
    (\partial_{t}^{2} + p^{2})\vec{A}_{1}^{p} = - i \mathcal{A}_0 \vec{\mathcal{P}}_{kp} e^{-ipt}\sum_{n=1}^{N}g_{a\gamma\gamma}^n m_n 
    \cos(m_n t + \theta_n)
\end{equation}
Our assumptions from the previous section still hold true with the additional information that $\theta_n$ also forms a similar set of identically distributed random variables
according to some distribution. In general, we expect such a distribution to be uniform in the range $[0,\ 2\pi]$ since there is no specific choice of phase. This means, for 
large $N$,
\begin{align}
    (\partial_{t}^{2} + p^{2})\vec{A}_{1}^{p} &= -\frac{i}{2\pi} N \mathcal{A}_0 \vec{\mathcal{P}}_{kp} e^{-ipt}\int_{g_{a\gamma\gamma}^L}^{g_{a\gamma\gamma}^M} g_{a\gamma\gamma} 
    \tilde{p}(g_{a\gamma\gamma}) \ d g_{a\gamma\gamma} \nonumber \\
    & \times \int_{m_L}^{m_M} \int_0^{2\pi} m_a \cos(m_a t+\theta) p(m_a)\ dm_a \ d\theta
\end{align}
Clearly, the $\theta$ integral vanishes - in other words, the random fluctuations in the phase average out to zero over the entire interval. Therefore, we consider a different
approach based on arguments similar to random walks. Instead of considering the average ``displacement'', we consider the average ``distance'' when treating random walks and in 
the same spirit, we move forward with the root mean square value here as well. In general, consider, 
\begin{equation}
    \sum_{n=1}^N g_{a\gamma\gamma}^n m_n \cos(m_n t + \theta_n) = \Re{\sum_{n=1}^N g_{a\gamma\gamma}^n m_n e^{i(m_n t+\theta_n)}}\ =\ \Re{Z_N} \
    =\ |Z_N| \cos[\Phi(t)]
\end{equation}
Therefore, the amplitude of our effective drive term is given by $|Z_n|$. Here, $\Phi(t)$ is some phase factor composed of some combination of $m_n$ and 
$\theta_n$. Now, 
\begin{equation}
    |Z_N|^2
    = \sum_{n=1}^N (g^n_{a\gamma\gamma} m_n)^2 + \frac{1}{2}\sum_{n\neq m}^N g_{a\gamma\gamma}^n g_{a\gamma\gamma}^m m_n m_m e^{i(\theta_n-\theta_m)} 
\end{equation}
The second term will average out to zero here. Thus, 
\begin{equation}
    |Z_N| \sim \sqrt{\sum_{n=1}^N (g^n_{a\gamma\gamma} m_n)^2}
\end{equation}
Based on this, we approximate eq. \eqref{incoherentoriginal} as, 
\begin{equation}
    (\partial_{t}^{2} + p^{2})\vec{A}_{1}^{p} = - i \mathcal{A}_0 \vec{\mathcal{P}}_{kp} e^{-ipt} 
    \sqrt{\sum_{n=1}^N (g_{a\gamma\gamma}^n m_n)^2} \cos(m_a^* t+\Theta)
\end{equation}
Where, 
\begin{equation}
    m^*_a = \int_{m_L}^{m_M} m_a p(m_a)\ d m_a
\end{equation}
Here, $\Theta$ is a global phase factor that we drop from our calculation. In other words, we replace the sum of random drive terms by a single, effective drive term with an amplitude as estimated above and a single mode of oscillation
depending on $m_a^*$. Crucially, our approximation will only be valid for drive terms that are closely spaced in frequency space - in 
other words, the string of cosine terms in our initial equation must be closely spaced. This implies that the ALP masses are closely spaced, i.e., 
the mass splitting parameter, defined as $m_M - m_L = 2\epsilon$ as before, is small (i.e., $\epsilon<<m_L$). We finally have,
\begin{equation}
    (\partial_{t}^{2} + p^{2})\vec{A}_{1}^{p} = -2 i \vec{D}_{k p}^N e^{-ipt} 
    \cos\left(m_a^* t\right)
\end{equation}
Here, 
\begin{equation}
    \vec{D}_{k p}^N = \frac{\mathcal{A}_0}{2} \vec{\mathcal{P}}_{kp} \sqrt{\sum_{n=1}^N (g_{a\gamma\gamma}^n m_n)^2}
\end{equation}
This equation is exactly similar to the one we encountered in the single ALP case with $m_a$ replaced by $m_a^*$. The resonant condition is given by, 
\begin{equation}
    p = \frac{1}{2}m_a^* = \frac{1}{2}\int_{m_L}^{m_M} m_a p(m_a)\ d m_a
\end{equation}
The resonant solution is given by, 
\begin{equation}
    \vec{A}(t, \vec{x}) = \vec{a}_{0} e^{i(\vec{k}\cdot\vec{x}-kt)} - 
    \frac{\mathcal{A}_0 t}{4} \sqrt{\sum_{n=1}^N (g_{a\gamma\gamma}^n m_n)^2} (\hat{k} \cross \vec{a}_{0}) e^{i(\vec{k}\cdot\vec{x}+kt)}
\end{equation}
The power carried by the echo wave is given by, 
\begin{equation}
    P_{N}  =  \left[\int_{g_{a\gamma\gamma}^L}^{g_{a\gamma\gamma}^M} g_{a\gamma\gamma}^2 \tilde{p}(g_{a\gamma\gamma})\ d g_{a\gamma\gamma}\right] \frac{\rho t}{16}
    \frac{dP_0}{d\nu}\Big|_{p=m_a^*/2}
\end{equation}
Therefore, we see that the incoherent case clearly removes the earlier $N$ dependence from the power. Another interesting point to note is that, unlike earlier, the mass distribution 
does not affect the power itself in any way. It simply influences the resonant condition, but the power itself is unaffected by what kind of distribution of masses we have. We can 
also express the power as, 
\begin{equation}
     P_{N}  =  f_g \Big[P_{N=1}\Big]_{p=m_a^*/2}^{g_{a\gamma\gamma}^M}
\end{equation}
Here, 
\begin{equation}
    f_g = \int_{g_{a\gamma\gamma}^L}^{g_{a\gamma\gamma}^M} \left(\frac{g_{a\gamma\gamma}}{g_{a\gamma\gamma}^M}\right)^2 \tilde{p}(g_{a\gamma\gamma})\ d g_{a\gamma\gamma}
    \leq 1
\end{equation}
The above integral is always less than unity for any smooth distribution - in fact, it is only equal to unity in the extreme case of a delta function (i.e., all couplings equal). This
essentially means that the incoherent case always leads to a weaker signal compared to the single ALP scenario, irrespective of the masses and couplings of the ALP fields. In fact, it can 
only approach the single ALP case as an extreme limit. We stress that this is a distinct feature of the multi-ALP framework where despite the signal
being weaker, the resonance condition is still influenced by the presence of multiple fields, rather than just one. Clearly, this has significant potential ramifications for 
interpreting experimental results whose conclusions will depend crucially on the underlying framework (single or multiple ALP) considered. We illustrate this point even more clearly 
in the following specific cases. 

\subsection{All masses equal}\label{subs1_2}
We proceed as we did in the coherent case. Let us first consider all masses and couplings equal. The resonant condition becomes, 
\begin{equation}
    p = \frac{1}{2}\lim_{\epsilon\rightarrow 0}\int_{m_a}^{m_a + 2\epsilon} m \delta(m-m_a)\ d m = \frac{1}{2}m_a
\end{equation}
As discussed earlier, the function $f(r_g)$ becomes unity in this case. The solution is given by,
\begin{equation}
    \vec{A}(t, \vec{x}) = \vec{a}_{0} e^{i(\vec{k}\cdot\vec{x}-kt)} - 
    \frac{g t}{4} (\hat{k} \cross \vec{a}_{0}) e^{i(\vec{k}\cdot\vec{x}+kt)}
\end{equation}
Where $g = g_{a\gamma\gamma}m_a \mathcal{A}_0$ as defined earlier. The power is, 
\begin{equation}
    P_{N}  =  P_{N=1}
\end{equation}
This is exactly the same as the single ALP case. The general feature of the incoherent case is the marked absence of the $N$ dependent amplification that we had earlier. Observationally, 
this means that we cannot differentiate between such a scenario and the single ALP case. However, this only happens in the extreme case when all of the masses and couplings are equal.
As an offshoot of this case, let us consider the situation where the couplings are still equal, but the masses are not. The resonant condition becomes, 
\begin{align}
    p = \frac{1}{2}\int_{m_L}^{m_L + 2\epsilon} m_a p(m_a)\ d m_a &= \epsilon p(m_L) m_L \Bigg[1 + \epsilon\left(\frac{1}{m_L} + \frac{p'(m_L)}{p(m_L)}\right)
    \nonumber \\
    &+\frac{2\epsilon^2}{3}\left(\frac{2p'(m_L)}{m_L p(m_L)}+\frac{p''(m_L)}{p(m_L)}\right)\Bigg] + \cdots \ =\ m_L f(\epsilon) a_1 + \cdots
\end{align}
Here, $f(\epsilon)$ and $a_1$ are as defined in eqs. \eqref{fdefine} and \eqref{adefine}. Now, the normalisation condition for the masses can be expressed as, 
\begin{equation}
    \int_{m_L}^{m_L + 2\epsilon} p(m)\ dm = 2 f(\epsilon) \left[1 + \frac{p'(m_L)}{p(m_L)} \epsilon + \frac{2 p''(m_L)}{3 p(m_L)}\epsilon^2 +\cdots\right] = 1
    \Rightarrow f(\epsilon) = \frac{1}{2} + \mathcal{O}(\epsilon^2)
\end{equation}
Therefore, the resonant condition is,
\begin{equation}
    p = \frac{1}{2} m_L + \mathcal{O}(\epsilon^2)
\end{equation}
While our resonance condition looked formidable initially, we have reduced it to a more tractable and experimentally relevant form. The power is simply, 
\begin{equation}
    P_{N}  =  P_{N=1}
\end{equation}
Therefore, both these cases reduce, in leading order, to the single ALP scenario.
\begin{figure}[h]
    \centering
    \includegraphics[width=0.85\textwidth]{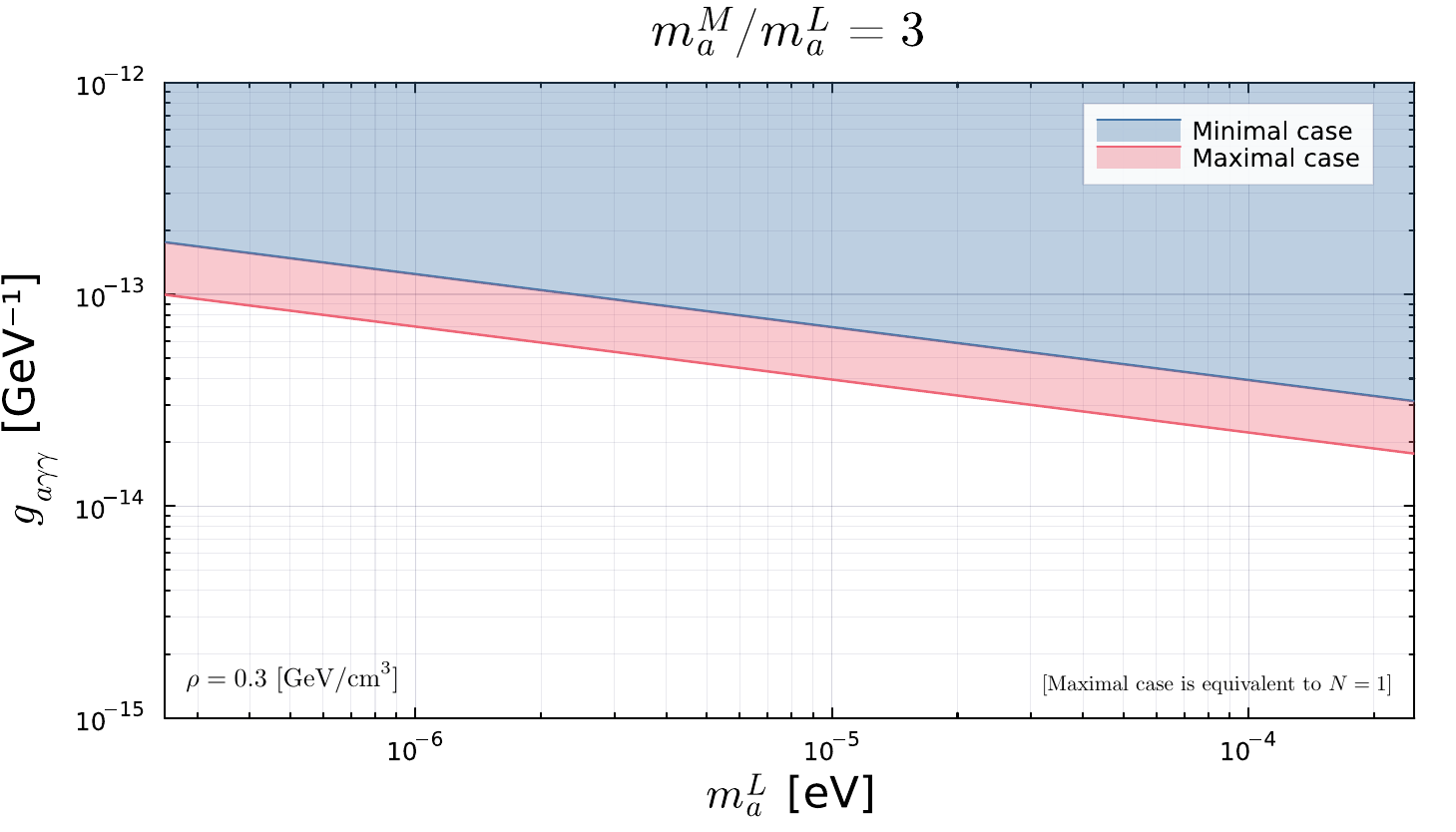}
    \caption{Sensitivity estimates for the minimal and maximal cases for the incoherent scenario with variable couplings in the isothermal model with $\rho = 0.3$ [GeV/cm$^3$].}
    \label{Fig_maxmincase}
\end{figure}

\subsection{Variable couplings and all masses equal}\label{subs2_2}
The next simplest case is when the masses are all equal, but the couplings are not. The resonant condition is simply $p = m_a/2$ as before. The solution is, 
\begin{equation}
    \vec{A}(t, \vec{x}) = \vec{a}_{0} e^{i(\vec{k}\cdot\vec{x}-kt)} - \frac{m_a \langle g_{a\gamma\gamma} \rangle \mathcal{A}_0}{4} t (\hat{k} \cross \vec{a}_{0}) 
    e^{i(\vec{k}\cdot\vec{x}+kt)}
\end{equation}
As with the previous case, the echo signal is modified this time as well - the deviation from the single ALP scenario depends on the exact choice of the coupling strengths.
The echo wave power can be written down as, 
\begin{equation}
    P_N =  f_g \Big[P_{N=1}\Big]_{g_{a\gamma\gamma}^M} = \langle g_{a\gamma\gamma} \rangle^2 \frac{t}{16} \rho \frac{dP_0}{d\nu}\Big|_{k=m_a/2}
\end{equation}
For uniform coupling distributions,
\begin{equation}
    P_N = \frac{1}{3}\left[\left(1 + \frac{g_{a\gamma\gamma}^L}{g_{a\gamma\gamma}^M}\right)^2 - \frac{g_{a\gamma\gamma}^L}{g_{a\gamma\gamma}^M}\right] \Big[P_{N=1}\Big]_{g_{a\gamma\gamma}^M}
\end{equation}
Once again, we have reduced the answer to two free ``effective'' couplings instead of $N$ different ones. Clearly, 
\begin{equation}
    \frac{1}{3}\Big[P_{N=1}\Big]_{g_{a\gamma\gamma}^M} \leq\ P_{N} =  f_g \Big[P_{N=1}\Big]_{g_{a\gamma\gamma}^M} \leq\ \Big[P_{N=1}\Big]_{g_{a\gamma\gamma}^M}
\end{equation}
As before, the projected sensitivities are weaker compared to the single ALP scenario. The exact sensitivities are dependent on the kind of coupling strength distribution in the underlying model. For uniform 
distributions, the maximal case corresponds to the single ALP scenario, while the minimal case corresponds to a third of it. The 
actual power, in any given situation, lies between this. This situation is depicted in figure \ref{Fig_maxmincase}. This result, while strikingly different from the coherent case, is 
as important and can potentially play a major role in interpreting experimental results.

\section{Discussion}\label{5}
In this manuscript, we have considered the case of echoes in the context of multiple ALPs interacting with photons. We have briefly revisited and cast the arguments presented in
earlier literature into a more tractable form and extended the formalism to cover the case of multiple ALPs. We have derived initial results for general distributions for both coherent and incoherent configurations and further,
considered various cases systematically - when the masses and couplings are all equal, when only the masses are equal and when only the couplings are equal. Our derivations have 
been presented in detail in the supplementary materials.\\

Specifically, we have shown that, in the coherent case with all ALP fields oscillating ``in phase'', the power scales as $N$ leading to a sharper amplification of the echo signal and 
strengthening the sensitivity estimates depending on how large $N$ is. This phenomenon is reminiscent of superradiance effects seen in other contexts, most notably discussed in
the seminal reference \cite{Dicke1954} - in our case, the coherent oscillations of the multiple ALP fields lead to a similar enhancement in the observable signal, except normalised by 
the fixed dark matter density in the universe. In case of small mass splittings between the ALPs, we have also shown that the mass splitting parameter itself 
provides an additional amplification. Such an amplification results in stronger sensitivities compared to the single ALP case. When discussing the incoherent case in section 
\ref{4}, we have shown how similar cases lead to a suppresion of the observable signal leading to weaker sensitivity estimates. Both the coherent and incoherent cases are, therefore, 
important for the proper interpretation of future experimental results. We stress that the coherent regime presents a benchmark scenario where the relative phases between the ALP fields
are negligible over the relevant interaction timescale ($t \sim 1/m_a$). This provides an absolute theoretical upper bound for any sort of multi-ALP signal. On the other hand, the
incoherent scenario shows that cosmological dephasing generically leads to echo signals typically weaker than single ALP power signals. This is in line with conclusions for 
multi-ALP DM in haloscope-based searches observed in reference \cite{Chadha-Day_2024}. The multiple ALP scenario, thus, presents certain distinct features that are not captured by single ALP
physics.\\

Further, we also discuss how our study, originally in the context of the QCD axion mass range, might also be 
applicable to other mass ranges, including the ultra-low frequency ranges that will be probed by future experiments, 
specifically NASA's FARSIDE and the DSL experiment in China. While sending out such low frequency photons into space in order to generate echoes might be impractical, our universe is 
already teeming with photons at these energies coming from various astrophysical sources including galactic emissions and even the Cosmic Microwave Background. Following the approach 
laid down in reference \cite{Echo7} and the groundwork in this manuscript, a detailed analysis of echoes generated from such galactic and extra-galactic sources at these ultra-light 
frequencies can be performed. Such studies could provide complementary bounds to existing data, for example, like those recently explored in references \cite{Taruya1,Taruya2,keck}.\\

More detailed and non-uniform probability distributions could still be effectively approximated by a smooth, uniform distribution if the mass 
range is quite narrow (which is similar to the case we considered) or if the variation in the masses is slow. In multiple ALP models, for example utilising the clockwork mechanism, 
the ALP masses are usually given by \cite{clockwork3}, 
\begin{equation}
    m^2_n = m^2 \left[q^2 + 1 - 2q \cos\left(\frac{n\pi}{N+1}\right) \right]
\end{equation}
The masses follow an approximately uniform distribution if $q$ is large enough; the relative spacing between the masses is $\mathcal{O}(1/q)$ rendering such a
treatment valid. The concept of an axion iceberg or a single broad signal which arises in such ALPs with very small mass splittings in the context of hadron colliders was recently 
explored in reference \cite{clockwork4}. Our treatment of the echoes in sections \ref{3} and \ref{4} would pertain to such situations and could provide complementary avenues for 
probing such beyond-standard-model scenarios. Some string theory models also generate ALP masses using a logarithmic distribution \cite{string2}, 
\begin{equation}
    P(m) \sim \ln\left(\frac{m}{M_{Pl}}\right)
\end{equation}
For ALPs with small mass splittings, such a distribution varies slowly enough to be approximated by a uniform distribution. While more detailed ALP setups would 
have more complicated mass spectra depending on the exact model, our results in the uniform case still hold qualitatively and are interesting due to the observational appeal and
the analytical tractability. Further, our results are general and can be applied to these cases properly, if required. Our study could provide a blueprint for more detailed 
model dependent calculations in this context and their implications which will be explored in a future work. 

\appendix

\section{Standard solution of a forced oscillator}\label{A1}

We start with, 
\begin{equation}
    (\partial_{t}^{2} + p^{2})\vec{A}_{1}^{p} = - ig\vec{\mathcal{P}}_{kp} e^{-ipt}\cos(m_a t)
\end{equation}
The homogenous equation is, 
\begin{equation}
    (\partial_{t}^{2} + p^{2})\vec{A}_{1}^{p} = 0
\end{equation}
This has two straightforward linearly independent solutions, 
\begin{equation}
    \vec{u}_1 = \vec{a} e^{ipt}, \ \vec{u}_2 = \vec{b} e^{-ipt}
\end{equation}
The Wronskian can be computed as, 
\begin{equation}
    \mathcal{W} = -2ip
\end{equation}
The particular solution can then be written as, 
\begin{equation}
    \vec{u}_p = \frac{i}{2p} e^{-ipt} \int_0^t d\xi\ \vec{G}(\xi)\ e^{ip\xi} - \frac{i}{2p} e^{ipt} \int_0^t d\xi\ \vec{G}(\xi)\ e^{-ip\xi}
\end{equation}
Here, $\vec{G}(t)$ is the inhomogenous term in our equation, 
\begin{equation}
    \vec{G}(t) = - ig\vec{\mathcal{P}}_{kp} e^{-ipt}\cos(m_a t)
\end{equation}
The first integral gives us, 
\begin{equation}
    \int_0^t d\xi\ \vec{G}(\xi)\ e^{ip\xi} = -ig\vec{\mathcal{P}}_{kp} \int_0^t d\xi\ \cos(m_a \xi) = -ig \vec{\mathcal{P}}_{kp} \frac{\sin(m_a t)}{m_a}     
\end{equation}
Similarly, the second integral gives us,
\begin{equation}
    \int_0^t d\xi\ \vec{G}(\xi)\ e^{-ip\xi} = -ig\vec{\mathcal{P}}_{kp} \int_0^t d\xi\ \cos(m_a \xi) e^{-2ip\xi} = 
    -\frac{g}{2} \vec{\mathcal{P}}_{kp} \left[\frac{e^{i(m_a - 2p)t}-1}{m_a-2p} - \frac{e^{-i(m_a + 2p)t}-1}{m_a+2p}\right]
\end{equation}
Or, 
\begin{equation}
    \int_0^t d\xi\ \vec{G}(\xi)\ e^{-ip\xi} =
    -\frac{g}{2} \vec{\mathcal{P}}_{kp} \left[\frac{e^{i(m_a - 2p)t}}{m_a-2p} - \frac{e^{-i(m_a + 2p)t}}{m_a+2p} - \frac{4p}{(m_a - 2p)(m_a+2p)}\right]
\end{equation}
Therefore, 
\begin{equation}
    \vec{u}_p = \frac{ig}{2m_a}\vec{\mathcal{P}}_{kp}\left[\frac{e^{i(m_a-p)t}}{m_a -2p} + \frac{e^{-i(m_a+p)t}}{m_a + 2p} -\frac{2m_a e^{ipt}}{(m_a-2p)(m_a+2p)}\right]
\end{equation}
Now, imposing the initial conditions $\vec{A}^p_1 (t=0) = 0,\ \Dot{A}^p_1 (t=0) = 0$, we get,
\begin{equation}
    \vec{a} = \vec{b} = 0
\end{equation}
Therefore, the complete solution is, 
\begin{equation}
    \vec{A}^p_1 =\frac{ig}{2m_a}\vec{\mathcal{P}}_{kp}\left[\frac{e^{i(m_a-p)t}}{m_a -2p} + \frac{e^{-i(m_a+p)t}}{m_a + 2p} -\frac{2m_a e^{ipt}}{(m_a-2p)(m_a+2p)}\right]
\end{equation}

\section{Validity of large $N$ approximation}\label{A3}

In this section, we discuss the validity of some of the statistical aspects used in the calculations. We have two sets of $N$ random variables, ${m_a^i}$ and 
$g_{a\gamma\gamma}^i$, with $i\in [1,N]$. We assume that each $g_{a\gamma\gamma}^i$ and $m_a^i$ is independently drawn from a continuous smooth distribution in a certain interval.
We further assume that each $g_{a\gamma\gamma}^i \in [g_{a\gamma\gamma}^L, g_{a\gamma\gamma}^M]$ and $m_a^i \in [m_L, m_M]\ \forall\ i \in [1, N]$. The probability distribution
for the coupling strengths is denoted by $\tilde{p}(g_{a\gamma\gamma})$ and the one for the ALP masses is $p(m_a)$. We also assume that the two distributions are independent of 
each other, i.e., uncorrelated. Now, the Law of Large Numbers can be expressed as the following 
theorem \cite{sheldonross}, 
\begin{theorem}
    Let $X_1,\ X_2,\ ...,\ X_n$ be a sequence of independent random variables having a common distribution, and let $E[X_i] = \mu$. Then, with probability $1$,
    \[ \frac{1}{n} \sum_{i=1}^{n}X_i \rightarrow \mu \ \text{as }n\rightarrow \infty \]
\end{theorem}
Based on this, let us look at the approximation we used in the main text, 
\begin{equation}
    \frac{1}{N}\sum_{n= 1}^{N} g_{a\gamma\gamma}^n m_n \cos(m_n t)= \int_{g_{a\gamma\gamma}^L}^{g_{a\gamma\gamma}^M}d g_{a\gamma\gamma}\ \tilde{p}(g_{a\gamma\gamma}) 
    \int_{m_L}^{m_M} d m_a\ p(m_a) \Big[g_{a\gamma\gamma} m_a \cos(m_a t)\Big]
\end{equation}
This is essentially the Law of Large Numbers written as, 
\begin{align}
     \frac{1}{N} \sum_{i=1}^{N} f(X_i) g(Y_i)\xrightarrow{\text{for large }N} \langle f(X_i)  g(Y_i)\rangle &= \int_{y_a}^{y_b} \int_{x_a}^{x_b} f(X_i) g(Y_i) \tilde{p}(X_i) p(Y_i) \ d X_i d Y_i
     \nonumber\\     
     &= \langle f(X_i) \rangle \langle g(Y_i)\rangle
\end{align}
Where we have implicitly assumed that the two distributions are independent of each other, i.e., uncorrelated. This occurs in the limit of very large $N$, ideally infinite. But our focus
in this section is to look at the validity of the above for finite $N$ - namely, how large do we need $N$ for the above to be roughly true? For this, let us first consider 
a situation where we only have a single set of random variables, 
\begin{equation}
    \frac{1}{N} \sum_{i=1}^{N} f(X_i) \rightarrow \langle f(X_i) \rangle = \int_a^b f(X_i) \tilde{p}(X_i) \ d X_i
\end{equation}
A simple and intuitive condition to quantify the fluctuations in $\langle f(X_i) \rangle$ is to impose, 
\begin{equation}
    \sqrt{\frac{1}{N}\text{Variance}\left[f(X_i)\right]} \equiv \frac{1}{\sqrt{N}}\sigma_f << \varepsilon^{1/2} \langle f(X_i) \rangle \equiv \varepsilon^{1/2} \mu_f
\end{equation} 
Where $\varepsilon$ is a real number that determines the precision of our approximation. Essentially, we are demanding that the fluctuations from the mean, relative to the number
of total ALP fields in the theory, is a small quantity - at least smaller than the mean. The parameter $\varepsilon$ quantifies the precision of our approximation. 
Satisfying the condition for $\varepsilon = 1$ is the standard, base case, while satisfying it for even smaller values of $\varepsilon$ hints at better validity. Our condition can 
also be expressed as, 
\begin{equation}
    N \varepsilon >> \left(\frac{\sigma_f}{\mu_f}\right)^2
\end{equation}
For our scenario, a straightforward generalisation gives, 
\begin{equation}
    N \varepsilon >> \left(\frac{\sigma_{fg}}{\mu_f \mu_g}\right)^2
\end{equation}
Where $\mu_g$ denotes the mean of $g(Y_i)$ with respect to its distribution and $\sigma_{fg}$ is the standard deviation of the quantity $f(X_i) g(Y_i)$, 
\begin{equation}
    \sigma_{fg}^2 = \langle [f(X_i) g(Y_i)]^2 \rangle - \langle f(X_i) g(Y_i) \rangle^2 = \langle f(X_i)^2\rangle \langle g(Y_i)^2 \rangle - \mu_f^2 \mu_g^2
\end{equation}
We can express this as, 
\begin{equation}
    \sigma_{fg}^2 = (\sigma_f^2 + \mu_f^2)(\sigma_g^2 + \mu_g^2) - \mu_f^2 \mu_g^2 = \sigma_f^2 \sigma_g^2 + \mu^2_f \sigma_g^2 + \mu^2_g \sigma_f^2
\end{equation}
Or,
\begin{equation}
    \left(\frac{\sigma_{fg}}{\mu_f \mu_g}\right)^2 = \left(\frac{\sigma_f \sigma_g}{\mu_f \mu_g}\right)^2 + \left(\frac{\sigma_f}{\mu_f}\right)^2 +
     \left(\frac{\sigma_g}{\mu_g}\right)^2
\end{equation}
Let us identify $\{Y_i\}$ with the ALP masses and $\{X_i\}$ with the coupling strengths. In the main text, we focused on scenarios where the ALP masses were concentrated in a narrow
range. Therefore, 
\begin{equation}
    \mu_g = \int_{y_a}^{y_a+\epsilon} dy\ g(y) p(y) = g(y_a) + g'(y_a)\int_{y_a}^{y_a+\epsilon} dy\ (y-y_a) p(y) + \cdots
\end{equation}
The second term in the equation above is $\mathcal{O}(\epsilon)$ and the terms following it get successively smaller. In fact, 
\begin{equation}
    \sigma_g^2 =  \int_{y_a}^{y_a+\epsilon} dy\ g(y)^2 p(y) - \mu_g^2\ \sim\ \mathcal{O}(\epsilon^2)
\end{equation}
This implies, 
\begin{equation}
    \left(\frac{\sigma_g}{\mu_g}\right)^2 \sim\ \mathcal{O}\left(\frac{\epsilon^2}{g(y_a)^2}\right) \rightarrow \text{very small}
\end{equation}
Since we are looking at a minimum bound for $N$, we can safely neglect these small additive terms. Therefore, we are left with, 
\begin{equation}
    \left(\frac{\sigma_{fg}}{\mu_f \mu_g}\right)^2 \approx \left(\frac{\sigma_f}{\mu_f}\right)^2
\end{equation}
It is not as simple to estimate the magnitude or place a bound on this quantity because, unlike the previous case, we do not have any further simplifying assumptions. In
general, it can be shown that, 
\begin{equation}
    \left(\frac{\sigma_f}{\mu_f}\right)^2 \leq\ \frac{(x_b-x_a)^2}{4 x_a x_b}\ \xrightarrow{x_b>>x_a}\ \mathcal{O}\left(1/r_g\right)
\end{equation}
Where, as in the main text, $r_g = g_{a\gamma\gamma}^L/g_{a\gamma\gamma}^M \equiv x_a/x_b$. Since $r_g$ is always less than unity, this provides a bound on $N$ that is finite, but
not necessarily small. However, this bound represents an extreme case based on the discrete two point distribution (where the probabilities are concentrated on the 
endpoints of the interval only) and, while true, does not provide a good description for the continuous and nice distributions that we are interested in. Unfortunately,
in order to place a better bound, one needs to make certain assumptions on the kind of distribution functions we have (which would be model dependent) - this can improve the above
result significantly. However, since such a treatment depends on the kind of models we have, we instead simply highlight that for simple uniform distributions, we have, 
\begin{equation}
    \left(\frac{\sigma_f}{\mu_f}\right)^2 = \frac{(x_b - x_a)^2}{3(x_b + x_a)^2}\ \xrightarrow{x_b>>x_a}\ \frac{1}{3} \ \Rightarrow \ N > \frac{1}{3\varepsilon}
\end{equation}
For $\varepsilon = 0.1$, we find that the condition holds true for $N>4$; this suggests that values of $N \sim \mathcal{O}(10)$ should be sufficient for our results to broadly hold
in practice. As noted earlier, it does not seem possible to formulate a more general bound applicable for all distributions without some further simplifying assumptions; nevertheless,
for physically relevant distributions, it is reasonable to expect that $N \sim \mathcal{O}(10)$ constitutes an adequate and reliable condition.

\section{Solution of the multi-ALP echo system}\label{A4}

\subsection{General solution}
We start with,
\begin{equation}
    (\partial_{t}^{2} + p^{2})\vec{A}_{1}^{p} = - 2 i N \vec{D}_{kp} e^{-ipt} Q(t)
\end{equation}
And, 
\begin{equation}
    Q(t) = \int_{m_L}^{m_M} m_a \cos(m_a t) p(m_a)\ dm_a
\end{equation}
In order to proceed further, we assume that the mass splitting, defined as $\epsilon = (m_M-m_L)/2$, is small. Then, 
\begin{equation}
    Q(t) = \int_{m_L}^{m_L + 2\epsilon} m_a \cos(m_a t) p(m_a)\ d m_a = \int_{m_L}^{m_L + 2\epsilon} h(m_a)\ d m_a = \bar{Q}(\epsilon, t) 
\end{equation}
Using the Leibniz rule, 
\begin{equation}
    \frac{d}{d\epsilon}\bar{Q}(\epsilon, t) = 2 h(m_L + 2\epsilon),\ \frac{d^2}{d\epsilon^2}\bar{Q}(\epsilon, t) = 4 \frac{d h(m_a+2\epsilon)}{d(m_a+2\epsilon)},\
    \frac{d^3}{d\epsilon^3}\bar{Q}(\epsilon, t) = 8 \frac{d^2 h(m_a+2\epsilon)}{d(m_a+2\epsilon)^2}
\end{equation}
Taylor expanding, we have,
\begin{multline}
    Q(t)\Big|_{m_M=m_L+2\epsilon} = \bar{Q}(0, t) + \epsilon\frac{d}{d\epsilon}\bar{Q}(0, t) + \frac{\epsilon^2}{2}\frac{d^2}{d\epsilon^2}\bar{Q}(0, t) + 
    \frac{\epsilon^3}{6}\frac{d^3}{d\epsilon^3}\bar{Q}(0, t)
    +\cdots \\= 2 h(m_L) \epsilon + 2 h'(m_L) \epsilon^2 + \frac{4}{3} h''(m_L) \epsilon^3 +\ldots
\end{multline}
Now, we have,
\begin{subequations}
\begin{align}
    h(m_L) &= m_L \cos(m_L t) p(m_L), \\ 
    h'(m_L) &= \cos(m_L t) p(m_L) - m_L t \sin(m_L t) p(m_L) + m_L \cos(m_L t) p'(m_L)
\end{align}
Further, 
\begin{align}
    h''(m_L) &= \left( -2t \sin(m_L t) - m_L t^2 \cos(m_L t) \right) p(m_L) \nonumber\\
&+ \left( 2 \cos(m_L t) - 2 m_L t \sin(m_L t) \right) p'(m_L)+ m_L \cos(m_L t) p''(m_L)
\end{align}
\end{subequations}
Or, to second order, 
\begin{align}
     Q(t)\Big|_{m_M=m_L+2\epsilon} &= 2 m_L \cos(m_L t) p(m_L) \left[1 + \epsilon\left(\frac{1}{m_L} + \frac{p'(m_L)}{p(m_L)}\right)+ 
    \frac{2\epsilon^2}{3}\left(\frac{2p'(m_L)}{m_L p(m_L)}+\frac{p''(m_L)}{p(m_L)}\right)\right]\epsilon \nonumber \\
    &- 2 m_L t \sin(m_L t) p(m_L)\left[1 + \frac{4\epsilon}{3}\left(\frac{1}{m_L} + \frac{p'(m_L)}{p(m_L)}\right)\right]\epsilon^2
\end{align}
Therefore, 
\begin{align}
    (\partial_{t}^{2} + p^{2})\vec{A}_{1}^{p} &= - 4 i N \vec{D}_{kp}^L f(\epsilon) e^{-ipt} \Bigg[\cos(m_L t)\left[1 + \frac{\epsilon}{m_L} + 
    \epsilon \frac{p'(m_L)}{p(m_L)} + \frac{2}{3}\epsilon^2 \left(\frac{2p'(m_L)}{m_L p(m_L)} + \frac{p''(m_L)}{p(m_L)}\right)\right]
    \nonumber \\ 
    &- t \sin(m_L t) \left[1 + \frac{4\epsilon}{3}\left(\frac{1}{m_L} + \frac{p'(m_L)}{p(m_L)}\right) \right]
    \epsilon- \frac{2}{3} \epsilon^2 t^2 \cos(m_L t)\Bigg]
\end{align}
Here, $f(\epsilon) = p(m_L)\epsilon$. Let us write this as, 
\begin{equation}
    (\partial_{t}^{2} + p^{2})\vec{A}_{1}^{p} = - 4 i N \vec{D}_{kp}^L f(\epsilon) e^{-ipt} \left[ a_1 \cos(m_L t) - a_2 \sin(m_L t)\epsilon t - \frac{2}{3}
    \epsilon^2 t^2 \cos(m_L t)\right]
\end{equation}
Here,
\begin{subequations}
\begin{align}
    a_1 &= \left[1 + \epsilon\left(\frac{1}{m_L} + \frac{p'(m_L)}{p(m_L)}\right) +\epsilon^2 b\right],\ b = \frac{2}{3}\left[\frac{p'(m_L)}{m_L p(m_L)} + \frac{p''(m_L)}{p(m_L)}\right], \\  
    a_2 &= \left[1 + \frac{4\epsilon}{3}\left(\frac{1}{m_L} + \frac{p'(m_L)}{p(m_L)}\right)\right]
\end{align}
\end{subequations}
Also, $\vec{D}_{kp}^L = \vec{D}_{kp} m_L$. Now, we need the solution to this. As outlined in appendix~\ref{A1}, our solution is given by,
\begin{equation}
    \vec{u}_p = \frac{i}{2p} e^{-ipt} \int_0^t d\xi\ \vec{G}(\xi)\ e^{ip\xi} - \frac{i}{2p} e^{ipt} \int_0^t d\xi\ \vec{G}(\xi)\ e^{-ip\xi}
\end{equation}
Where,
\begin{equation}
    \vec{G}(t) = - 4 i N \vec{D}_{kp}^L f(\epsilon) e^{-ipt} \left[ a_1 \cos(m_L t) - a_2 \sin(m_L t)\epsilon t  - \frac{2}{3}\epsilon^2 t^2 \cos(m_L t)\right]
\end{equation}
The first integral can be computed in a straightforward fashion, 
\begin{align}
    \int_0^t d\xi\ \vec{G}(\xi)\ e^{ip\xi} &= 
    - 4 i N \vec{D}_{kp}^L f(\epsilon) \int_0^t d\xi\ \left[ a_1 \cos(m_L \xi) - a_2 \sin(m_L \xi)\epsilon \xi - \frac{2}{3}\epsilon^2 \xi^2 \cos(m_L\xi)\right]
    \nonumber\\ 
    &= - 4 i N \vec{D}_{kp}^L f(\epsilon)\Bigg[\frac{a_1}{m_L}\sin(m_L t) + \frac{\epsilon a_2}{(m_L)^2}\left[m_L t\cos(m_L t) - \sin(m_L t)\right]
    \nonumber\\
    &- \frac{2\epsilon^2}{3(m_L)^3} \left[2m_L t\cos(m_L t) + ((m_L t)^2 -2)\sin(m_L t)\right] \Bigg]\nonumber\\
    &\equiv - 4 i N \vec{D}_{kp}^L f(\epsilon) \mathcal{Q}_1 (p, t)
\end{align}
Similarly, let us take a look at the second integral,
\begin{equation}
    \int_0^t d\xi\ \vec{G}(\xi)\ e^{-ip\xi} = 
    - 4 i N \vec{D}_{kp}^L f(\epsilon) \int_0^t d\xi\ \left[a_1 \cos(m_L \xi) - a_2 \sin(m_L \xi)\epsilon \xi - \frac{2}{3}\epsilon^2 \xi^2 \cos(m_L\xi)\right]e^{-2ip\xi}
\end{equation}
Let us define (for $\alpha \in \mathbb{R}$ and $\alpha \neq 0$), 
\begin{equation}
    \mathcal{J}_n^\pm (\alpha, t) = \int_0^t d\xi\ \xi^n e^{\pm i \alpha \xi} 
    = \mp \frac{i}{\alpha} t^n e^{\pm i \alpha t} \pm \frac{i}{\alpha} n \int_0^t d\xi\ \xi^{n-1} e^{\pm i \alpha \xi}
    = \mp \frac{i}{\alpha} t^n e^{\pm i \alpha t} \pm \frac{i}{\alpha} n \mathcal{J}_{n-1}^\pm (\alpha, t)
\end{equation}
Here, $n\geq0$. The base case is,
\begin{equation}
    \mathcal{J}_0^\pm (\alpha, t) = \int_0^t d\xi\ e^{\pm i \alpha \xi} = \mp \frac{i}{\alpha} \left[e^{\pm i \alpha t}-1\right]
\end{equation}
Then, 
\begin{equation}
    \mathcal{J}_1^\pm (\alpha, t) = \mp \frac{i}{\alpha} t e^{\pm i \alpha t} \pm \frac{i}{\alpha} \mathcal{J}_{0}^\pm (\alpha, t) = 
    \mp \frac{i}{\alpha} t e^{\pm i \alpha t} \pm \frac{i}{\alpha} \left(\mp \frac{i}{\alpha} e^{\pm i \alpha t} \pm \frac{i}{\alpha}\right)
    = \mp \frac{i}{\alpha} t e^{\pm i \alpha t} + \frac{1}{\alpha^2} \left[e^{\pm i \alpha t}-1\right]
\end{equation}
And, 
\begin{equation}
    \mathcal{J}_2^\pm (\alpha, t) =\mp \frac{i}{\alpha} t^2 e^{\pm i \alpha t} + \frac{2}{\alpha^2} t e^{\pm i \alpha t} \pm \frac{2i}{\alpha^3} \left[e^{\pm i \alpha t}-1\right]
\end{equation}
Using these definitions, we can write,
\begin{align}
    \int_0^t d\xi\ \vec{G}(\xi)\ e^{-ip\xi} &= 
    - 4 i N \vec{D}_{kp}^L f(\epsilon) \int_0^t d\xi\ \left[a_1 \cos(m_L \xi) - a_2 \sin(m_L \xi)\epsilon \xi - \frac{2}{3}\epsilon^2 \xi^2 \cos(m_L\xi)\right]e^{-2ip\xi} 
    \nonumber \\
    &= -2 i N \vec{D}_{kp}^L f(\epsilon)\Bigg[a_1\Big[\mathcal{J}^+_0(m_L-2p, t) + \mathcal{J}^-_0(m_L+2p, t)\Big] +i a_2 \epsilon
     \Big[\mathcal{J}^+_1(m_L-2p, t) 
     \nonumber\\ 
     &- \mathcal{J}^-_1(m_L+2p, t)\Big]- \frac{2}{3}\epsilon^2 \Big[\mathcal{J}^+_2(m_L-2p, t) + \mathcal{J}^-_2(m_L+2p, t)\Big] \Bigg]
     \nonumber\\
    &\equiv - 4 i N \vec{D}_{kp}^L f(\epsilon) \mathcal{Q}_2 (p, t)
\end{align}
Therefore, the total solution is, 
\begin{equation}
    \vec{A}^p_1 = \frac{2}{p} N \vec{D}_{kp}^L f(\epsilon) \left[e^{-ipt} \mathcal{Q}_1 (p, t) - e^{ipt} \mathcal{Q}_2 (p, t)\right]
\end{equation}
With, 
\begin{align}
    \mathcal{Q}_1 (p, t) &= \frac{a_1}{m_L}\sin(m_L t) + \epsilon \frac{a_2}{(m_L)^2}\left[m_L t\cos(m_L t) - \sin(m_L t)\right] 
    \nonumber \\
    &-\frac{2\epsilon^2}{3(m_L)^3} \left[2m_L t\cos(m_L t) + ((m_L t)^2 -2)\sin(m_L t)\right] 
\end{align}
And,
\begin{align}
    \mathcal{Q}_2 (p, t) &= \frac{1}{2}\Bigg[a_1\left[\mathcal{J}^+_0(m_L-2p, t) + \mathcal{J}^-_0(m_L+2p, t)\right] +i a_2 \epsilon
     \left[\mathcal{J}^+_1(m_L-2p, t) - \mathcal{J}^-_1(m_L+2p, t)\right]\nonumber
     \\ &- \frac{2}{3}\epsilon^2 \left[\mathcal{J}^+_2(m_L-2p, t) + \mathcal{J}^-_2(m_L+2p, t)\right] \Bigg]
\end{align}

\subsection{Resonant response}
Let us consider the system at resonance, i.e., $p = m_L/2 + \delta$. It is quite clear the resonant term is $\mathcal{Q}_2(p, t)$ - specifically the terms with 
($m_L-2p$) in their arguments. We consider only the dominant contributions, 
\begin{equation}
    \text{At resonance: }\mathcal{Q}_2 (p, t) \approx \frac{1}{2}\Big[a_1 \mathcal{J}^+_0(m_L-2p, t) +i\epsilon a_2\mathcal{J}^+_1(m_L-2p, t)- 
    \frac{2}{3}\epsilon^2 \mathcal{J}^+_2(m_L-2p, t)\Big]
\end{equation}
As per our earlier definitions, at resonance, the individual terms are,
\begin{equation}
    \mathcal{J}^+_0(m_L-2p, t) = i\frac{e^{-2i\delta t}-1}{2\delta},\ \mathcal{J}^+_1(m_L-2p, t) = 
    it\frac{e^{ -2i  \delta t}}{2\delta} + \frac{e^{-2i \delta t}-1}{4 \delta^2}
\end{equation}
\begin{equation}
    \mathcal{J}^+_2(m_L-2p, t) = 
    i t^2 \frac{e^{-2i \delta t}}{2\delta} + t\frac{e^{ -2i \delta t}}{2 \delta^2}  -i \frac{e^{-2i \delta t}-1}{4 \delta^3}
\end{equation}
Therefore, at resonance,
\begin{align}
    \mathcal{Q}_2 (p, t) &= \frac{1}{2}\left[i\frac{e^{-2i\delta t}-1}{2\delta}\left(a_1 + \frac{\epsilon}{2\delta} a_2+\frac{\epsilon^2}{3\delta^2}\right)
    - t e^{-2i\delta t}\left(\frac{\epsilon}{2\delta} a_2+\frac{\epsilon^2}{3\delta^2}\right)- i\frac{\epsilon^2}{3\delta}t^2 e^{-2i \delta t} \right]\nonumber
   \\ &= \frac{1}{2}\Bigg[i\frac{e^{-2i\delta t}-1}{2\delta}\left[1 + \frac{\epsilon}{m_L} + \epsilon \frac{p'(m_L)}{p(m_L)} + \epsilon^2 b+ \frac{\epsilon}{2\delta} 
    \left[1 + \frac{4\epsilon}{3}\left(\frac{1}{m_L} + \frac{p'(m_L)}{p(m_L)}\right) \right]+\frac{\epsilon^2}{3\delta^2}\right]\nonumber
    \\&- \left[\left[1 + \frac{4\epsilon}{3}\left(\frac{1}{m_L} + \frac{p'(m_L)}{p(m_L)}\right) \right] \frac{\epsilon}{2\delta}+\frac{\epsilon^2}{3\delta^2} \right]t 
    e^{-2i\delta t}- i\frac{\epsilon^2}{3\delta}t^2 e^{-2i \delta t}\Bigg]
\end{align}
Let us expand the sums in orders of $\epsilon$. We have,
\begin{equation}
   \mathcal{O}(\epsilon^0):\ i\frac{e^{-2i\delta t}-1}{2\delta} =\sum_{n=1}^{\infty} \frac{(-t)^n}{n!} i^{n+1} (2\delta)^{n-1}
\end{equation}
And,
\begin{align}
    \mathcal{O}(\epsilon^1)&:\ i\frac{e^{-2i\delta t}-1}{2\delta}\left(\frac{1}{m_L}+ \frac{1}{2\delta} + \frac{p'(m_L)}{p(m_L)}\right) - \frac{t}{2\delta}e^{-2i\delta t}\nonumber
    \\ &= \left[\frac{1}{m_L}+\frac{p'(m_L)}{p(m_L)}\right]\sum_{n=1}^{\infty} \frac{(-t)^n}{n!} i^{n+1} (2\delta)^{n-1} - \sum_{n=1}^{\infty}\frac{n}{(n+1)!}(-i)^n t^{n+1}(2\delta)^{n-1} 
\end{align}
Finally, 
\begin{align}
    \mathcal{O}(\epsilon^2)&:\ \left[\frac{2}{3 \delta}\left[\frac{1}{m_L} + \frac{p'(m_L)}{p(m_L)}\right]+\frac{1}{3\delta^2}\right]\left[i\frac{e^{-2i\delta t}-1}{2\delta}- 
    t e^{-2i\delta t}\right]+ib\frac{e^{-2i\delta t}-1}{2\delta}-\frac{i}{3\delta}t^2 e^{-2i \delta t} \nonumber
   \\ &= b\sum_{n=1}^{\infty} \frac{(-t)^n}{n!} i^{n+1} (2\delta)^{n-1} - \frac{4}{3}\left[\frac{1}{m_L} + \frac{p'(m_L)}{p(m_L)}\right]\sum_{n=1}^{\infty}
   \frac{n}{(n+1)!}(-i)^n t^{n+1}(2\delta)^{n-1} \nonumber 
   \\ &-\frac{1}{3} \sum_{n=1}^{\infty}\frac{n}{(n+2)n!}(-2)^n i^{n+1} t^{n+2}\delta^{n-1}
\end{align}
Putting everything together, we get, 
\begin{align}
    \mathcal{Q}_2 (p, t)
    &= \frac{1}{2}\sum_{n=1}^{\infty} \frac{(-it)^n}{n!}(2\delta)^{n-1}\Bigg[i\left(1 + \frac{\epsilon}{m_L}+\epsilon \frac{p'(m_L)}{p(m_L)}+b\right) 
    - \frac{n}{n+1}\left(1 + \frac{4\epsilon}{3m_L}+\epsilon \frac{4p'(m_L)}{3p(m_L)}\right)\epsilon t \nonumber \\ &- \frac{2in}{3(n+2)}\epsilon^2 t^2\Bigg]
     = \frac{1}{2}\sum_{n=1}^{\infty} \frac{(-it)^n}{n!}(2\delta)^{n-1}\left[i a_1
    - \frac{n}{n+1}a_2 \epsilon t - \frac{2in}{3(n+2)}\epsilon^2 t^2\right]
\end{align}
For $\delta\rightarrow0$, we have,
\begin{equation}
    \mathcal{Q}_2 (p, t) = \frac{t}{2}\left[a_1 + \frac{i}{2} a_2 \epsilon t - \frac{2}{9}\epsilon^2 t^2\right]
\end{equation}

\subsection{Real space solution}
Let us now consider the behaviour of our previous solution in the resonant limit,
\begin{equation}
    \vec{A}^p_1 = \frac{2}{p} N \vec{D}_{kp}^L f(\epsilon) \left[e^{-ipt} \mathcal{Q}_1 (p, t) - e^{ipt} \mathcal{Q}_2 (p, t)\right]
\end{equation}
Now, at each $\epsilon$ order, we can see that the terms of $\mathcal{Q}_2(p, t)$ are dominant since they are polynomials in $t$, as opposed to oscillatory terms. Thus, 
\begin{equation}
    \text{At resonance: }\vec{A}^p_1 = -\frac{t}{p} N \vec{D}_{k p}^L f(\epsilon) e^{ipt}\left[a_1 + \frac{i}{2} a_2 \epsilon t- \frac{2}{9}\epsilon^2 t^2\right]
\end{equation}
Thus, the solution in real space is, 
\begin{equation}
    \vec{A}(t, \vec{x}) = \vec{a}_{0} e^{i(\vec{k}\cdot\vec{x}-kt)} - 
    N\frac{\tilde{g}t}{2} (\hat{k} \cross \vec{a}_{0})  \left[a_1 + \frac{i}{2} a_2 \epsilon t- \frac{2}{9}\epsilon^2 t^2\right] e^{i(\vec{k}\cdot\vec{x}+kt)}
\end{equation}
Where, 
\begin{equation}
    \tilde{g} = m_L  f(\epsilon) \mathcal{A}_0 \int_{g_{a\gamma\gamma}^L}^{g_{a\gamma\gamma}^M} g_{a\gamma\gamma} \tilde{p}(g_{a\gamma\gamma}) \ d g_{a\gamma\gamma}
\end{equation}

\subsection{Power calculation}
In order to estimate the power carried by the echo wave, we start from, 
\begin{equation}
    \mathcal{Q}_2 (p, t) = \frac{1}{2}\sum_{n=1}^{\infty} \frac{(-it)^n}{n!}(2\delta)^{n-1}\left[i a_1
    - \frac{n}{n+1}a_2 \epsilon t- \frac{2in}{3(n+2)}\epsilon^2 t^2\right]
\end{equation}
We need to calculate the magnitude of the above to compute the power. Let us define,
\begin{align}
    \mathcal{Y}_1 (x) = -\frac{\epsilon t}{4\delta}\sum_{n=1}^{\infty} \frac{(-it)^n}{n!}(2\delta)^{n} \frac{x^n}{n+1}
    &= -\frac{\epsilon t}{4\delta} \int_0^1 dy\ \sum_{n=1}^{\infty} \frac{(-it)^n}{n!}(2\delta)^{n} y^n x^n \nonumber\\
    &= -\frac{\epsilon t}{4\delta} \int_0^1 dy\ \left( e^{-2i\delta xyt}-1\right)
\end{align}
And, 
\begin{align}
    \mathcal{Y}_2(x) = -i\frac{\epsilon^2 t^2}{6\delta}\sum_{n=1}^{\infty}\frac{(-it)^n}{n!}(2\delta)^{n}\frac{x^n}{n+2}
    &= -i\frac{\epsilon^2 t^2}{6\delta}\int_0^1 dy\ \sum_{n=1}^{\infty}\frac{(-2i\delta x t)^n}{n!} y^{n+1} \nonumber\\
    &= -i\frac{\epsilon^2 t^2}{6\delta}\int_0^1 dy\ y\left(e^{-2i\delta xyt}-1\right)
\end{align}
Then,
\begin{equation}
    \mathcal{Q}_2 (p, t) 
     = a_1 \frac{i}{4\delta}(e^{-2i\delta t}-1) + a_2 \frac{\partial \mathcal{Y}_1}{\partial x}\Bigg|_{x=1} + \frac{\partial \mathcal{Y}_2}{\partial x}\Bigg|_{x=1}
     = \frac{t}{2}
    \int_0^1 dy\ e^{-2i\delta yt}\Big(a_1 + i a_2 \epsilon t  y - \frac{2}{3}\epsilon^2 t^2 y^2\Big)
\end{equation}
The magnitude is, 
\begin{align}
    |\mathcal{Q}_2(p, t)|^2 &= \frac{t^2}{4} \int_0^1 dy \int_0^1 dx\  e^{2i\delta t (x-y)}
    \Big[a_1 + i a_2 \epsilon t  y- \frac{2}{3}\epsilon^2 t^2 y^2\Big]\Big[a_1 - i a_2 \epsilon t  x - \frac{2}{3}\epsilon^2 t^2 x^2\Big]\nonumber
    \\ &= \frac{t^2}{4} \int_0^1 dy \int_0^1 dx\  e^{2i\delta t (x-y)}\Big[a_1^2 + \left(a_2 \epsilon t\right)^2 xy + i a_1 a_2 \epsilon t (y-x)
    - \frac{2}{3}a_1 \epsilon^2 t^2 \left(x^2+y^2\right) \nonumber\\
    &- \frac{2i}{3}a_2 \epsilon^3 t^3 x y (x-y) + \mathcal{O}(\epsilon^4)\Big]
\end{align}
Clearly, if $t$ is large, the exponential will oscillate rapidly - therefore, the major contribution will always be from the region where $x,\ y$ are close to each other. 
Thus, let us define new variables,
\begin{equation}
    \alpha = x+y,\ \beta = x-y
\end{equation}
The Jacobian is simply $1/2$. Therefore,
\begin{align}
    |\mathcal{Q}_2(p, t)|^2  &=\frac{t^2}{8} \int_\mathcal{D} d\alpha\ d\beta\  e^{2i\delta t \beta}
    \Bigg[a_1^2 + i a_1 a_2 \epsilon t \beta 
    + \left[\left(\frac{a_2^2}{4}-\frac{a_1}{3}\right)\alpha^2 - \left(\frac{a_2^2}{4}+\frac{a_1}{3}\right)\beta^2\right](\epsilon t)^2\nonumber
    \\ &- \frac{i}{6}a_2 \epsilon^3 t^3 \beta (\alpha^2 - \beta^2)\Bigg]
\end{align}
$\mathcal{D}$ denotes the new region of integration after the change of variables,
\begin{equation}
    \int_0^1 dx\ \int_0^1 dy\ \rightarrow \left(\int_{0}^{1} \int_{-\alpha}^{\alpha} +\int_{1}^{2} \int_{\alpha-2}^{2-\alpha}\right) \,d\beta\,d\alpha
\end{equation}
Let us evaluate this integral term by term. We switch the $\beta$ limits to cover the entire real line because the major contribution to the integral only comes from 
the small $\beta$ region. We also ignore the fourth order term. We have,
\begin{equation}
    \int_{-\infty}^{\infty} e^{2i\delta t \beta}\,d\beta = 2\pi \delta(2\delta t) = \frac{\pi}{t}\delta(\delta)
\end{equation}
Further, 
\begin{equation}
    \int_{-\infty}^{\infty} \beta e^{2i\delta t \beta}\,d\beta \rightarrow 0 \ \text{as }\delta\rightarrow 0
\end{equation}
The other odd term in $\beta$ also vanishes similarly. Next, 
\begin{equation}
    \left(\int_{0}^{1} \int_{-\alpha}^{\alpha} +\int_{1}^{2} \int_{\alpha-2}^{2-\alpha}\right)\beta^2 e^{2i\delta t \beta}\,d\beta\,d\alpha \xrightarrow{\lim \delta\rightarrow 0}
   2\pi \delta(2\delta t)\left(\int_{0}^{1} \int_{-\alpha}^{\alpha} +\int_{1}^{2} \int_{\alpha-2}^{2-\alpha}\right)\beta^2 \,d\beta\,d\alpha = \frac{\pi}{3t}\delta(\delta)
\end{equation}
Putting everything together,
\begin{align}
    |\mathcal{Q}_2(p, t)|^2
    &= \frac{t^2}{8} \int_\mathcal{D} d\alpha\ d\beta\  e^{2i\delta t \beta}\left[a_1^2 + i a_1 a_2 \epsilon t \beta 
    + \left[\left(\frac{a_2^2}{4}-\frac{a_1}{3}\right)\alpha^2 - \left(\frac{a_2^2}{4}+\frac{a_1}{3}\right)\beta^2\right](\epsilon t)^2\right]\nonumber
    \\ &= \frac{\pi t}{4}\delta(\delta)\left[a_1^2 +\frac{7}{24}\left(a_2^2-\frac{12}{7}a_1\right)\epsilon^2 t^2\right]
\end{align}
Therefore, the power is, 
\begin{align}
    P_{N}^\epsilon &= \int dp\ |\vec{A}^p_1|^2 = \int dp\ \frac{4N^2}{p^2} |\vec{D}_{k p}^L f(\epsilon)|^2 |\mathcal{Q}_2 (p, t)|^2\nonumber
    \\ &= \frac{4\pi t \mathcal{A}_0^2 N^2}{16} [m_L f(\epsilon)]^2 
    \left[\int_{g_{a\gamma\gamma}^L}^{g_{a\gamma\gamma}^M} g_{a\gamma\gamma} \tilde{p}(g_{a\gamma\gamma}) \ d g_{a\gamma\gamma}\right]^2   
    \left[a_1^2 +\frac{7}{24}\left(a_2^2-\frac{12}{7}a_1\right)\epsilon^2 t^2\right]\frac{dP_0}{dp}\Bigg|_{p=m_L/2}
\end{align}
Now, the DM density is given by, 
\begin{equation}
    \rho = \frac{1}{2}\sum_{n=1}^N (m_n)^2 \mathcal{A}_0^2 = \frac{1}{2}N \mathcal{A}_0^2 \int_{m_L}^{m_L + 2\epsilon} m_a^2\ p(m_a) \ d m_a
    \equiv \frac{1}{2}N \mathcal{A}_0^2 I(\epsilon)
\end{equation}
Where, 
\begin{equation}
    I(\epsilon) = \int_{m_L}^{m_L + 2\epsilon} m_a^2\ p(m_a) \ d m_a = \int_{m_L}^{m_L + 2\epsilon} h(m_a)\ d m_a 
\end{equation}
Taylor expanding and proceeding as before, we have, 
\begin{align}
    I(\epsilon)&= 2 (m_L)^2 p(m_L)\epsilon \left[1 + \epsilon\left(\frac{1}{m_L} + \frac{p'(m_L)}{p(m_L)}\right)+ 
    \frac{2\epsilon^2}{3}\left(\frac{2p'(m_L)}{m_L p(m_L)}+\frac{p''(m_L)}{p(m_L)}\right)\right]\nonumber
    \\ &+ 2 m_L p(m_L)\left[1 + \frac{4\epsilon}{3}\left(\frac{1}{m_L} + \frac{p'(m_L)}{p(m_L)}\right)\right]\epsilon^2 + \cdots 
\end{align}
Thus, 
\begin{equation}
    \rho = N \mathcal{A}_0^2 (m_L)^2 f(\epsilon)\left[a_1 + \frac{\epsilon}{m_L}a_2\right]
\end{equation}
Therefore, our final expression for the power is, 
\begin{equation}
    P_{N}^\epsilon  
    = 2 N \mathcal{Z}(\epsilon, t) P_{N=1}\Big|_{g_{a\gamma\gamma}^M}
\end{equation}
Where,
\begin{equation}
    \mathcal{Z}(\epsilon, t) = f(\epsilon)
    \left[\int_{g_{a\gamma\gamma}^L}^{g_{a\gamma\gamma}^M} \frac{g_{a\gamma\gamma}}{g_{a\gamma\gamma}^M} \tilde{p}(g_{a\gamma\gamma}) \ d g_{a\gamma\gamma}\right]^2
    \left[a_1 + \frac{\epsilon}{m_L}a_2\right]^{-1}\left[a_1^2 +\frac{7}{24}\left(a_2^2-\frac{12}{7}a_1\right)\epsilon^2 t^2\right]
\end{equation}
Here, $P_{N=1}$ refers to the power in the original, single ALP case.

\acknowledgments

SH would like to thank Jayanta K. Bhattacharjee, Anuraj Chatterjee and Tanmoy Kumar for several helpful discussions. SH would also like to acknowledge the KVPY fellowship provided 
by the Department of Science and Technology (DST), Government of India.

\bibliographystyle{JHEP}
\bibliography{Bibliography}

\end{document}